\documentclass[usenatbib,a4paper]{mn2e}
\usepackage{amsmath}
\usepackage{amssymb}
\usepackage{amsfonts}
\usepackage {multirow}
\usepackage {graphicx}
\usepackage {natbib}
\usepackage{txfonts}
\usepackage{graphicx}
\usepackage{soul}
\usepackage{float}
\usepackage{threeparttable}
\usepackage[bottom]{footmisc}
\usepackage{threeparttable}
\usepackage[hyperindex,breaklinks=true, colorlinks, citecolor=blue]{hyperref}
\usepackage{tablefootnote}
\usepackage{eurosym}
\usepackage{ulem}
\usepackage[usenames,dvipsnames]{xcolor}
\usepackage{booktabs}
\usepackage[mathscr]{euscript}
\DeclareSymbolFont{rsfs}{U}{rsfs}{m}{n}
\DeclareSymbolFontAlphabet{\mathscrsfs}{rsfs}
\usepackage{url}
\usepackage{multicol}
\usepackage{multirow}
\usepackage{makecell}
\usepackage{pifont}
\usepackage{slashed}
\usepackage{verbatim}
\usepackage{aas_macros}
\usepackage{ae,aecompl}
\usepackage{mathtools}
\usepackage[english]{babel}
\usepackage{float} 
\usepackage{url}

\def\setsymbol#1#2{\expandafter\def\csname #1\endcsname{#2}}
\def\getsymbol#1{\csname #1\endcsname}

\newbox\tablebox    \newdimen\tablewidth
\def\leaderfil{\leaders\hbox to 5pt{\hss.\hss}\hfil}
\def\tablenote#1 #2\par{\begingroup \parindent=0.8em
    \abovedisplayshortskip=0pt\belowdisplayshortskip=0pt
    \noindent
    $$\hss\vbox{\hsize\tablewidth \hangindent=\parindent \hangafter=1 \noindent
    \hbox to \parindent{$^#1$\hss}\strut#2\strut\par}\hss$$
    \endgroup}

\def\L2{\ifmmode L_2\else $L_2$\fi}

\def\DeltaT{\ifmmode \Delta T\else $\Delta T$\fi}
\def\deltat{\ifmmode \Delta t\else $\Delta t$\fi}
\def\fknee{\ifmmode f_{\rm knee}\else $f_{\rm knee}$\fi}
\def\Fmax{\ifmmode F_{\rm max}\else $F_{\rm max}$\fi}
\def\solar{\ifmmode{\rm M}_{\mathord\odot}\else${\rm M}_{\mathord\odot}$\fi}
\def\Msolar{\ifmmode{\rm M}_{\mathord\odot}\else${\rm M}_{\mathord\odot}$\fi}
\def\Lsolar{\ifmmode{\rm L}_{\mathord\odot}\else${\rm L}_{\mathord\odot}$\fi}
\def\inv{\ifmmode^{-1}\else$^{-1}$\fi}
\def\mo{\ifmmode^{-1}\else$^{-1}$\fi}
\def\sup#1{\ifmmode ^{\rm #1}\else $^{\rm #1}$\fi}
\def\expo#1{\ifmmode \times 10^{#1}\else $\times 10^{#1}$\fi}
\def\,{\thinspace}
\def\lsim{\mathrel{\raise .4ex\hbox{\rlap{$<$}\lower 1.2ex\hbox{$\sim$}}}}
\def\gsim{\mathrel{\raise .4ex\hbox{\rlap{$>$}\lower 1.2ex\hbox{$\sim$}}}}

\def\simprop{\mathrel{\raise .4ex\hbox{\rlap{$\propto$}\lower 1.2ex\hbox{$\sim$}}}}
\def\deg{\ifmmode^\circ\else$^\circ$\fi}
\def\pdeg{\ifmmode $\setbox0=\hbox{$^{\circ}$}\rlap{\hskip.11\wd0 .}$^{\circ}
          \else \setbox0=\hbox{$^{\circ}$}\rlap{\hskip.11\wd0 .}$^{\circ}$\fi}
\def\arcs{\ifmmode {^{\scriptstyle\prime\prime}}
          \else $^{\scriptstyle\prime\prime}$\fi}
\def\arcm{\ifmmode {^{\scriptstyle\prime}}
          \else $^{\scriptstyle\prime}$\fi}
\newdimen\sa  \newdimen\sb
\def\parcs{\sa=.07em \sb=.03em
     \ifmmode \hbox{\rlap{.}}^{\scriptstyle\prime\kern -\sb\prime}\hbox{\kern -\sa}
     \else \rlap{.}$^{\scriptstyle\prime\kern -\sb\prime}$\kern -\sa\fi}
\def\parcm{\sa=.08em \sb=.03em
     \ifmmode \hbox{\rlap{.}\kern\sa}^{\scriptstyle\prime}\hbox{\kern-\sb}
     \else \rlap{.}\kern\sa$^{\scriptstyle\prime}$\kern-\sb\fi}
\def\ra[#1 #2 #3.#4]{#1\sup{h}#2\sup{m}#3\sup{s}\llap.#4}
\def\dec[#1 #2 #3.#4]{#1\deg#2\arcm#3\arcs\llap.#4}
\def\deco[#1 #2 #3]{#1\deg#2\arcm#3\arcs}
\def\rra[#1 #2]{#1\sup{h}#2\sup{m}}

\def\dots{\relax\ifmmode \ldots\else $\ldots$\fi}
\def\WHzsr{\ifmmode $W\,Hz\mo\,sr\mo$\else W\,Hz\mo\,sr\mo\fi}
\def\mHz{\ifmmode $\,mHz$\else \,mHz\fi}
\def\GHz{\ifmmode $\,GHz$\else \,GHz\fi}
\def\mKs{\ifmmode $\,mK\,s$^{1/2}\else \,mK\,s$^{1/2}$\fi}
\def\muKs{\ifmmode \,\mu$K\,s$^{1/2}\else \,$\mu$K\,s$^{1/2}$\fi}
\def\muKRJs{\ifmmode \,\mu$K$_{\rm RJ}$\,s$^{1/2}\else \,$\mu$K$_{\rm RJ}$\,s$^{1/2}$\fi}
\def\muKHz{\ifmmode \,\mu$K\,Hz$^{-1/2}\else \,$\mu$K\,Hz$^{-1/2}$\fi}
\def\MJysr{\ifmmode \,$MJy\,sr\mo$\else \,MJy\,sr\mo\fi}
\def\MJysrmK{\ifmmode \,$MJy\,sr\mo$\,mK$_{\rm CMB}\mo\else \,MJy\,sr\mo\,mK$_{\rm CMB}\mo$\fi}
\def\microns{\ifmmode \,\mu$m$\else \,$\mu$m\fi}

\def\muK{\ifmmode \,\mu$K$\else \,$\mu$\hbox{K}\fi}
\def\microK{\ifmmode \,\mu$K$\else \,$\mu$\hbox{K}\fi}
\def\muW{\ifmmode \,\mu$W$\else \,$\mu$\hbox{W}\fi}
\def\kms{\ifmmode $\,km\,s$^{-1}\else \,km\,s$^{-1}$\fi}
\def\kmsMpc{\ifmmode $\,\kms\,Mpc\mo$\else \,\kms\,Mpc\mo\fi}
\providecommand{\sorthelp}[1]{}

\DeclareMathAlphabet{\mathsc}{OT1}{cmr}{m}{sc}
\newcommand{\healpix}{\ensuremath{\tt HEALPix}}
\newcommand{\NILC}{\ensuremath{\tt NILC}}

\newcommand{\COMMANDER}{\ensuremath{\tt Commander}}

\newcommand{\COrE}{\ensuremath{\tt COrE }}
\newcommand{\camb}{\ensuremath{\tt CAMB}}
\newcommand{\planck}{{\it Planck}}

\newcommand{\lenspix}{\ensuremath{\tt LensPix }}
\newcommand{\GNILC}{\ensuremath{\tt GNILC }}

\newcommand{\hi}{\ensuremath{\mathsc {Hi}}}
\newcommand{\GNILCn}{\ensuremath{\tt GNILC-dust }}
\newcommand{\GINES}{\ensuremath{\tt MKD-dust }}
\newcommand{\TDdust}{\ensuremath{\tt TD-dust }}
\newcommand{\galprop}{\ensuremath{\tt GALPROP }}
\newcommand{\smica}{\ensuremath{\tt SMICA }}
\newcommand{\powerlaw}{\ensuremath{\tt Power-law }}
\newcommand{\curvedpowerlaw}{\ensuremath{\tt Curved-power-law}}
\newcommand{\cmark}{\ding{51}}
\newcommand{\xmark}{\ding{55}}
\newcommand{\namaster}{\ensuremath{\tt NaMaster }}

\newcommand{\ECHO}{\ensuremath{\tt ECHO}}
\newcommand{\pico}{\ensuremath{\tt PICO}}
\newcommand{\prism}{\ensuremath{\tt PRISM}}
\newcommand{\cmbbh}{{\tt CMB-Bh$\overline{a}$rat}}

\newcommand{\Nside}{\ensuremath{N_{\rm side}}}
\def\parcm{$^{\scriptstyle\prime}$}
\def\GHz{\ifmmode $\,GHz$\else \,GHz\fi}
\def\muK{\ifmmode \,\mu$K$\else \,$\mu$\hbox{K}\fi}
\def\MJysrmK{\ifmmode \,$MJy\,sr\mo$\,mK$_{\rm CMB}\mo\else \,MJy\,sr\mo\,mK$_{\rm CMB}\mo$\fi}

\newcommand\subsubsubsubsection{\@startsection{subparagraph}{5}{\z@}{-2.5ex\@plus -1ex \@minus -.25ex}{1.25ex \@plus .25ex}{\normalfont\normalsize\bfseries}}
\makeatother

\def\reff@jnl#1{{\rm#1\/}}

\def\solphys{\reff@jnl{Solar~Phys.}}    
\def\sovast{\reff@jnl{Soviet~Ast.}}     
\def\ssr{\reff@jnl{Space~Sci.Rev.}}    
\def\zap{\reff@jnl{ZAp}}                
\def\nat{\reff@jnl{Nature}}             
\def\aj{\reff@jnl{AJ}}                  
\def\araa{\reff@jnl{ARA\&A}}            
\def\apj{\reff@jnl{ApJ}}                
\def\aapr{\reff@jnl{A\&A~Rev.}}         
\def\aaps{\reff@jnl{A\&AS}}             
\def\azh{\reff@jnl{AZh}}                        
\def\baas{\reff@jnl{BAAS}}              
\def\jcap{\reff@jnl{JCAP}}              
\def\jrasc{\reff@jnl{JRASC}}            
\def\memras{\reff@jnl{MmRAS}}           
\def\mnras{\reff@jnl{MNRAS}}            
\def\pra{\reff@jnl{Phys.Rev.A}}         
\def\prb{\reff@jnl{Phys.Rev.B}}         
\def\prc{\reff@jnl{Phys.Rev.C}}         
\def\prd{\reff@jnl{Phys.Rev.D}}         
\def\prl{\reff@jnl{Phys.Rev.Lett}}      
\def\pasp{\reff@jnl{PASP}}              
\def\pasj{\reff@jnl{PASJ}}              
\def\qjras{\reff@jnl{QJRAS}}            
\def\skytel{\reff@jnl{S\&T}}            
\def\procspie{\reff@jnl{Proceedings of the SPIE}}             
\def\apjl{\reff@jnl{ApJ}}               
\def\apjs{\reff@jnl{ApJS}}              
\def\ao{\reff@jnl{Appl.Optics}}         
\def\apss{\reff@jnl{Ap\&SS}}            
\def\aap{\reff@jnl{A\&A}}   

\newcommand{\daa}[1]{{\textcolor{magenta}{#1}}}

\title[ECHO]{$B$-mode forecast of CMB-Bh$\overline{a}$rat}

\author[Adak et al.]{Debabrata Adak$^{1}$\thanks{E-mail: debabrata@iucaa.in},
Aparajita Sen$^{2}$\thanks{aparajita15@iisertvm.ac.in},
Soumen Basak$^{2}$,
Jacques Delabrouille$ ^{3,4,5}$,\newauthor   Tuhin Ghosh$^{6}$, Aditya Rotti$^{7}$, Gin\'es Mart\'{i}nez-Solaeche$^{8}$, Tarun Souradeep$^{10,9,1}$
\\
$^{1}$ Inter University Centre for Astronomy and Astrophysics, Post Bag 4, Ganeshkhind, Pune-411007, India\\
$^{2}$ School of Physics, Indian Institute of Science Education and Research Thiruvananthapuram,Maruthamala PO, Vithura,\\ Thiruvananthapuram 695551, Kerala, India\\
$^{3}$ APC, CNRS/IN2P3, Universit\'{e} Paris Diderot, 10, rue Alice Domon et L\'eonie Duquet, 75205 Paris Cedex 13, France, \\
$^{4}$ IRFU, CEA, Universit\'{e} Paris Saclay, 91191 Gif-sur-Yvette, France\\
$^{5}$ Department of Astronomy, School of Physical Sciences, University of Science and Technology of China, Hefei, Anhui 230026, China.\\
$^{6}$ School of Physical Sciences, National Institute of Science Education and Research, HBNI, Jatni 752050, Odisha, India\\
$^{7}$ Jodrell Bank Centre for Astrophysics, School of Physics and Astronomy, The University of Manchester, \\Oxford Road, Manchester, M13 9PL, U.K.\\
$^{8}$ Instituto de Astrofi\'{s}ica de Andaluci\'{a} (CSIC), PO Box 3004, 18080 Granada, Spain\\
$^{9}$ Indian Institute of Science Education and Research,  Dr. Homi  Bhabha Road,  Pashan, Pune  411008, Maharashtra, India\\
$^{10}$ Raman Research Institute, C. V. Raman Avenue, Bengaluru, Karnataka 560080, India\\
}
\date{January 2021}

\begin{document}
\date{\vspace{-.6mm}{Accepted  -- Received }}
\maketitle

\begin{abstract}
Exploring Cosmic History and Origins (\ECHO), popularly known as  `\cmbbh', is a space mission that has been proposed to the Indian Space Research Organisation (ISRO) for the scientific exploitation of the Cosmic Microwave Background (CMB) at the next level of precision and accuracy. The quest for the CMB polarization $B$-mode signals, generated by inflationary gravitational waves in the very early universe, is one of the key scientific goals of its experimental design. This work studies the potential of the proposed \ECHO\ instrumental configuration to detect the target tensor-to-scalar ratio $r \sim 10^{-3}$ at $3\sigma$ significance level, which covers the predictions of a large class of inflationary models. We investigate the performance of two different component separation pipelines, \NILC\ and  \COMMANDER, for the measurement of $r$ in presence of different physically motivated models of astrophysical foregrounds. For a simplistic foreground model (only polarized dust and synchrotron), both component separation pipelines can achieve the desired sensitivity of \ECHO, i.e. $\sigma (r =0) \sim (0.4 - 0.7)\times 10^{-3}$. \NILC\ performs better than \COMMANDER\ in terms of bias on recovered $r$ for complex spectral models (power-law and curved power-law) of the synchrotron emission and complex dust models (dust decorrelation). 
Assuming 84\% delensing, we can achieve an improvement of $\sigma (r = 0)$ by approximately 50\% as compared to the results obtained for the same configuration without any lensing correction.
\end{abstract}
%

%
\section{Introduction}
\label{sec:intro}
The Cosmic Microwave Background (CMB) provides direct information about the earliest moments of our universe, and our current understanding about the origin and the subsequent cosmological evolution of the universe is heavily based on measurements of this relic radiation. Over the last couple of decades, a huge amount of information has been accumulated about the temperature and polarization anisotropies of CMB. Ground-based, Balloon-borne and Satellite missions {\citep{PEPI:2011, Polarbear:2014,aiola2020atacama,ACT:2020, SPT:2020,gualtieri2018spider,aghanim2020planck}} have measured the CMB anisotropies at various angular scales and frequencies. All these observations have successfully converged into the $\Lambda$CDM model, or concordance cosmology (See \cite{BULL201656} for review):
a spatially flat expanding universe composed of matter, baryons and vacuum energy with a nearly scale-invariant spectrum of primordial fluctuations. The global properties of the universe are characterized by the density of baryons, cold dark matter and dark energy. 
Primordial perturbations of the space-time metric, i.e. the initial conditions for our observable universe, are thought to originate from a phase of
cosmic inflation {\citep{STAROBINSKY:1980,1980ApJ...241L..59K,Guth:1980,10.1093/mnras/195.3.467,linde1982new,PhysRevLett.48.1220}.}
During this epoch, the universe is stretched so violently that tiny quantum fluctuations are imprinted on cosmological distance scales. 
These fluctuations grow over the time due to gravitational instability to give all the large scale structure that we observe today.

A generic prediction of inflationary models is the generation of a stochastic background of gravitational waves, the ripples in the fabric of space-time, which gives rise to a faint but distinctive twisting pattern in the polarization of the CMB, known as $B$-modes \citep{Seljak:1997, Hu_and_white:1997, Kamionkowski:1998}. While the amplitude of the $B$-mode signal is unknown, the shape of the spectrum is robustly predicted by theory. The primordial $B$-mode power spectrum has a characteristic double-humped shape, the first bump on large angular angular scales ($\ell$ < 10) produced at reionization epoch and the second on degree scales produced during electron-photon decoupling around the time of recombination. 
 The $B$-mode power at degree angular scales is related to the amount of primordial gravitational waves emitted during inflation, and is quantified by the tensor-to-scalar ratio $r$. The amplitude of this background is determined by the energy scale of inflation \citep{Knox_Turner:1993}, which can widely vary among different inflationary models. 
 CMB $B$-mode polarization hence give us a mean to investigate the physics of the inflation. This makes the search for the $B$-mode polarization crucially linked to a probe of the inflationary scenario.

\planck\ is considered the definitive mission about CMB temperature anisotropies at angular scales larger than $5$ arcmin. However, the sensitivity of \planck\ was insufficient to extract all the information carried by CMB polarization.
For a definitive measurement, the sensitivity to polarization signals needs to be improved at least by one or two orders of
magnitude. 

CMB polarization $B$-modes are known to be much fainter that CMB temperature anisotropies and polarization $E$-modes. Hence, their detection constitutes a major technological challenge. However, recent developments in detector technology have made this exciting prospect possible. Although the signal still remains elusive, the latest observations of the sky by the Planck satellite has put an upper limit on the tensor-to-scalar ratio $r_{0.05} < 0.056$ at 95\% confidence level. This upper limit is further tightened by combining with the BICEP2/Keck Array BK18 data to obtain $r_{0.05} < 0.036$ \citep{tristram2021planck_BK18}. The upper bound on $r$ combined with the the measurement of the primordial spectral index $n_s$ have been able to rule out a large number of 
 inflationary models. The simplest slow-roll models that naturally explain the observed departure of $n_s$ from unity predict $r > 0.001$. The inflation models such as Starobinsky $R^2$,  $\alpha$-attractor and D-brane inflation are strongly favored while some class of models such as natural inflation, low-scale SUSY (supersymmetry) have been ruled out with significance of more than $2\sigma$ level \citep{akrami2020planck}.  A detection of the $B$-modes or even achieving a tighter upper bound of $r \sim 10^{-3}$ will give us a deeper understanding of the inflationary physics and its energy scale that would shed light on the mechanism that created the primordial perturbations.

The importance of this scientific goal has triggered the scientific community to build highly sensitive instruments and plan for new ground-based and space-based projects. Among the experiments that are currently ongoing or in preparation, ground based experiments are expected to be limited by their restricted frequency and sky coverage, while the LiteBIRD \citep{Hazumi:2019} space mission, owing to its limited angular resolution ($\sim$30\parcm), will be limited by confusion between primordial and lensing polarization $B$-modes. To overcome these limitations, other experiments have been proposed, such as \COrE\ \citep{2011arXiv1102.2181T,2018COREmission}, \prism\ \citep{PRISM},  \pico\ \citep{PICO:2019}, and more recently an ambitious spectro-polarimetric survey of the microwave sky \citep{Delabrouille1:2019}. Exploring Cosmic History and Origin (\ECHO)\footnote{https://cmb-bharat.in/} is the latest addition to this list with similar objectives. 

\ECHO\ is a proposal for a next generation space mission for near-ultimate measurements of the CMB polarization and discovery of global CMB spectral distortions. The proposal is primarily inspired from the concept of the \COrE\ mission and borrows many of its most appealing design aspects \citep{2018COREmission,core_instrument:2018}, and it is under consideration to the Indian Space Research Organization (ISRO)\footnote{\url{https://www.isro.gov.in/}}. 
The objective of this paper is to forecast the potential of this mission to achieve the main goal, the measurement of tensor-to-scalar ratio. Similar forecasting constraints on $r$ have been done for many proposed or ongoing CMB missions \citep{2016MNRAS4582032R,2016JCAP03052E,David:2017, CORE-B:2018, PICO:2019, CMBS4_2022}.

The paper is arranged as follows. Section~\ref{sec:echo} describes in detail the objective of \ECHO\ space mission: foreground cleaning challenge and the specification of the instrument. This is followed by a discussion of the simulation of the microwave sky for different foreground models and component separation methods used for the analysis in Section~\ref{sec: sim} and Section~\ref{sec:method} respectively. The main results of the analysis are summarized in Section~\ref{sec:results}. Finally, we conclude in Section~\ref{sec:Conclusions}.

%

%
\section{ECHO}
\label{sec:echo} 

\subsection{Sensitivity objectives and requirements}
\label{sec:sensitivity_requirement}
\ECHO\ aims to detect and characterize the primordial gravitational waves  by measuring $B$-modes of CMB polarization signal at large angular scales. A quantitative goal for the full success of \ECHO\ is to achieve a 68\% confidence level on measurement of the tensor-to-scalar ratio of the order of $ 10^{-3}$.
Achieving this science goal requires:
\begin{itemize}
   
    \item Low detector noise and measurement errors: State of the art detectors built using current technology are limited by the fundamental photon noise, and so the sensitivity of individual detectors cannot be improved much. Hence, the only way to improve the sensitivity of the instruments, is to build large focal plane instruments with thousands of detectors, which, in effect, make many simultaneous measurements.  
    In addition to this detectors, one must cope with the thermal emission of the telescope and the endowment around it. This limits our choice to place the instruments because only space instruments are expected to be cold enough to avoid such local emission.
    A very subtle shift or rotation of a detector, or glint of light from out of the field could mimic a spurious signal. The design of the instruments, the program of mapping and observation, the experimental characterization and calibration of the experiment will all need to be carried out with an unparalleled level of precision so that the instrumental systematic effects must be controlled to level well below the sensitivity of the instrument.
    
    \item Removal of  foreground contamination: The capability to distinguish CMB $B$-mode polarization from contamination by galactic and extragalactic astrophysical emissions is determined by multitude of astrophysical signals, frequency range and number of bands the mission has across the microwave and sub-millimeter range. The foreground and cosmological signals are separable using their spectral signatures. Although CMB is same at all frequency bands (in thermodynamic unit), the electromagnetic spectrum of the astrophysical components are different across frequencies. The CMB $B$-mode signal is expected to be very weak and buried in the astrophysical foregrounds (we discuss this issue in more detail in Section \ref{sec: fore}). Thus, to characterize astrophysical foreground emission and efficiently subtract them from CMB, we need large frequency coverage.
    
    \item Removal of contamination due to lensing: At large angular scales ($\ell$ < 150), the $B$-modes transformed from $E$-modes due to gravitational lensing resemble as white noise of approximately constant amplitude of 5 $\mu$K.arcmin \citep{PhysRevD.69.043005}. 
    The lensing $B$-modes and primordial $B$-modes are comparable at $\ell$ = 80 for tensor-to-scalar ration $r \approx$ 0.01. The lensing $B$-modes has dominant contribution to uncertainties of $r$ measurement if noise level is lower than the lensing signal of 5\,$\mu$K.arcmin. Therefore, to reach the target level of tensor-to-scalar ratio of the order of $10^{-3}$,  lensing correction is necessary. This requires an angular resolution of a few arcmin at observation frequencies in the $100-200$\,GHz range (and below if possible).
\end{itemize}

To take these requirements into account, the \ECHO\ satellite will scan the microwave sky in temperature and polarization in 20 frequency bands ranging from few tens of \GHz\ to THz. The satellite will be placed in orbit around Sun-Earth second Lagrange (L2) point. The observations will be carried out for 4 years in the anti-solar direction, pointing away from contaminating radiation from the Sun, the Earth and the Moon.
The instrument will measure the CMB polarization field down to angular scales of a few arcminutes with a sensitivity of 1-2 $\mu$K.arcmin. 

\begin{figure}
\includegraphics[width=\linewidth]{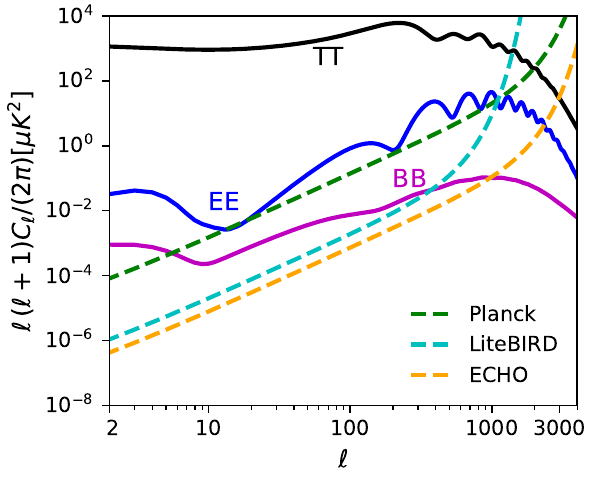} \par
\caption{Theoretical CMB angular power spectra of temperature $T$, $E$- and $B$-mode polarization anisotropies as a function of multipole moments for the best-fit $\Lambda$CDM parameters derived by \planck\  \citep{planck-VI:2018} with $r$ = 0.06.
The instrument noise power spectrum of polarization (dash lines) plotted for different generations of space based CMB surveys, viz. \planck, LiteBIRD and \ECHO.} 
\label{fig:echo_noise_spectra}
\end{figure}

Table~\ref{table2} presents the instrument specification of \ECHO. \ECHO\ is inspired by the \COrE\ concept and borrows many of its most appealing design aspects \citep{2018COREmission,core_instrument:2018}. It improves the \COrE\ concept by extending the frequency coverage at both high and low end of frequency coverage while also enhancing the sensitivity roughly by a factor of $\sqrt{2}$. \ECHO\ scanning strategy will cover the full sky with multiple visits to the same sky location over a period of 4 years of observations. Fig.~\ref{fig:echo_noise_spectra} compares the target \ECHO\ noise power spectra polarization with those of different generations of space based CMB surveys (existing or planned). 

\begin{figure}
\includegraphics[width=\columnwidth]{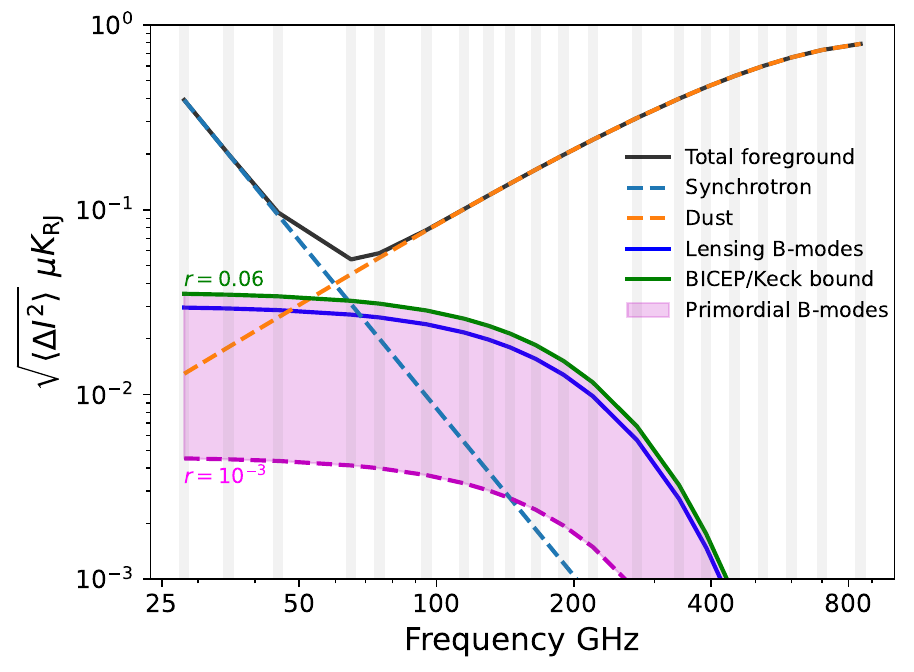} \par
\caption{This figure depicts the root mean square (r.m.s) fluctuations associated with the synchrotron, dust and the CMB $B$-mode signal. The  solid black line represents the total foreground. 
The magenta band denotes the CMB for the range of $r \in [0.001, 0.06]$. 
The vertical gray bands mark the \ECHO\ channels.} 
\label{fig:sky_spectra}
\end{figure}

\ECHO\ is designed to allow for maximum control over foregrounds. Fig.~\ref{fig:sky_spectra} shows that three \ECHO\ channels below $50$ GHz are dominated by synchrotron emission while the $14$ channels above $100$ GHz are dominated by dust emission. The high frequency channels of the instrument, particularly the frequency channels from 220 to 900 \GHz, are designed to remove the dust contamination in CMB and measure the Cosmic Infrared Background (CIB) anisotropies. On the other hand the low frequency channels are designed to mitigate the contamination of CMB polarization measurements by Galactic synchrotron emission. The channels between $50$ and $200$ GHz can be thought of as the CMB channels, that are expected to drive the detection of the primordial $B$-mode signal. In essence, \ECHO\ configuration aims to observe the polarized microwave sky with $10-30$ times better sensitivity than \planck\ and will more than double the frequency coverage.

\begin{table}
\caption{\ECHO\ instrument specification as proposed in the \cmbbh\ proposal.}
\label{table2}
   \begin{tabular}{ p{2.cm}  p{2.cm}   p{3.2cm}  }
   \hline\\
   Frequency & Beam FWHM  & $Q$ $\&$ $U$ noise r.m.s\\
   (GHz)&(arcmin)&($\mu$K.arcmin)\\
   \hline
   \hline
   28&39.9&16.5\\
   35&31.9&13.3\\
   45&24.8&11.9\\
   65&17.1&8.9\\
   75&14.91&5.1\\
   95&11.7&4.6\\
   115&9.72&3.1\\
   130&8.59&3.1\\
   145&7.70&2.4\\
   165&6.77&2.5\\
   190&5.88&2.8\\
   220&5.08&3.3\\
   275&4.06&6.3\\
   340&3.28&11.4\\
   390&2.86&21.9\\
   450&2.48&43.4\\
   520&2.14&102.0\\
   600&1.86&288.0\\
   700&1.59&1122.0\\
   850&1.31&9550.0\\
 \hline\\
   \end{tabular}
\end{table}
%
%

%
%

\subsection{The foreground issue} 
\label{sec: fore}
The weakness of the CMB polarization signal  makes it heavily contaminated by foreground emission from both the galactic and extragalactic sources. The extragalactic foregrounds such as point sources are approximately white noise,  whereas the galactic synchrotron and dust emission are expected to have a significant polarization at large angular scales. The minimum polarized foreground contamination is roughly corresponding to $r_{\rm FG} = $  $0.06$ to $1.0$ in the frequency range $60-100$ \GHz\ \citep{Krachmalnicoff:2016}. In fact, depending on the amplitude of $r$ and scaling of foreground components, the signal could be buried nearly two orders of magnitude below the foreground contamination as shown in Fig.~\ref{fig:sky_spectra}. Given the current upper limits on $r$, there is no regions of the sky which are dominated by the primordial $B$-mode signal. Therefore a robust measurement of primordial $B$-modes will require foreground removal to anticipated levels. 
In order to achieve the target of \ECHO, residual in the recovered CMB  maps should be controlled below $ 10^{-5} \mu K^{2}$.

Unlike the intensity foregrounds, the polarized foreground sky is relatively less complex and, dominated primarily by synchrotron and dust emissions. The diffuse synchrotron emission is generated from gyrating relativistic cosmic ray leptons in the galactic magnetic field. It is modeled by the density of cosmic ray leptons, their spectral properties, the galactic magnetic field \citep{Fauvet:2011, Orlando:2013}. The thermal dust emissions from elongated grains aligned in galactic magnetic field \citep{Lazarian:2007} is the dominant galactic emission at frequencies greater than 100 \GHz. The emission law of the thermal dust can be fitted with a modified black-body (MBB) \citep{planck-x:2016}. In addition to these two main astrophysical emissions, anomalous microwave emission (AME) from spinning dust grains with low polarization fraction, may be a relevant component at $10-60$\,\GHz\ \citep{ Leitch:1997, Oliveira-Costa:2004, Finkbeiner:2004}. Extragalactic objects also emit polarized light that have varied spectral energy distribution and polarization properties \citep{Saikia:1988} having a median polarization of 2\% \citep{2017MNRAS4692401B,2018A&A618A29T,2018ApJ85885P}. However, the impact of these sources is relatively small at large angular scales. 

In recent years, quite a large number of component separation methods have been developed. Some of them are already implemented on high quality cosmological data sets such as \planck. While these methods primarily aim at cleaning CMB maps, they differ from each other in their domain of implementation. While the primary goal of \ECHO\ is to study the CMB polarization maps, a great interest lies in a deeper understanding of the polarization properties of astrophysical emission processes too. Hence, we approach the CMB cleaning in two different ways: treating CMB as one of the many physical signal we want to recover together with synchrotron and thermal dust emissions, or focusing on isolating the CMB regardless of the physical mechanisms responsible for the contaminating emissions.

The component separation pipelines considered in this work are the \COMMANDER\ \citep{Eriksen:2004, Eriksen:2008} and the Needlet Internal Linear Combination (\NILC, \citealt{J.Delabrouille:2009,Basak_and_Delabrouille:2012,Basak:2013}). In the \COMMANDER\ pipeline, the foreground parameters are fitted through Bayesian parameter estimation technique operating on real (pixel) space. 
\NILC\  is an implementation of Internal Linear Combination (ILC,  \citealt{1996MNRAS.281.1297T}) of the frequency channels under consideration with minimum error variance, on a frame of spherical wavelets called needlets \citep{doi:10.1137/040614359}. This method allows localized filtering in both pixel space and harmonic space. This particular implementation of ILC has the advantage that the weights used to combine the sky can vary with a position on the sky and also with an angular scale.

For $r \sim 10^{-3}$, the cosmic variance errors associated with lensing $B$-modes will limit the statistical significance with which primordial $B$-mode signal can be measured. Therefore, to improve the sensitivity of $r$ measurement, the lensing contribution must be reduced exploiting the delensing techniques with the aid of CMB lensing map, CIB map, external galaxy survey data sets \citep{Sherwin_Schmittfull:2015, Manzotti:2017, Manzotti:2017a, Baleato:2021}. \ECHO\ is design to provide high resolution $E$-mode and CIB maps after applying the component separation techniques to the total frequency maps which will help in lensing correction.

\section{Sky modeling}
\label{sec: sim}
In this section, we describe the models of each of the sky components (CMB, thermal dust, synchrotron, AME and point sources) used in our analysis.

Since the prime focus of this work is to study the $B$-modes of CMB polarization at large angular scales, we consider only the polarized emission of the foregrounds and do not include the secondary anisotropies such as Sunyaev-Zel'dovich effect, patchy reionization effect \citep{Smith:2017, SuvodipMukherjee:2019} and any extragalactic diffuse emission e.g. CIB in our analysis. We integrate over top-hat bandpass profile around central frequencies in simulating the maps.

\subsection{CMB}
\label{sec:cmb}
We use \planck\ 2018 best-fit $\Lambda$CDM model \citep{planck-VI:2018} for tensor-to-scalar ratio $r$ = 0 to model the CMB sky.
 The theoretical angular spectra of CMB and lensing potential are obtained using \camb\footnote{\url{https://camb.info/}}. The lensed CMB maps are generated from the theory spectra using the \lenspix\footnote{\url{http://cosmologist.info/lenspix/}} package. For \COrE\ mission, it has been investigated that reduction of the lensing $B$-mode power is possible by 70 \% using CMB lensing and CIB maps \citep{CORE-B:2018}.  
Since \ECHO\ instrument design is motivated by \COrE\ concept, we are optimistic to achieve similar or plausibly better delensing in future. Therefore, in addition to fully lensed CMB maps, we also consider 84 \% delensed CMB maps in our work. These maps are generated from the weighted linear combination of the unlensed and lensed CMB maps as 
\begin{equation}
\text{Total} \,\,\,CMB = \sqrt{A_{L}} \times \mathrm{lensed}\,\,\mathrm{CMB} + (1-\sqrt{A_{L}}) \times \mathrm{unlensed}\,\,\mathrm{CMB} \ ,
\end{equation}
where $A_{L}$ is the lensing amplitude. $A_{L}=1$ for no delensing case and $A_{L}=0.16$ for 84\% delensing case.


\subsection{Thermal dust model}
\label{sec:dust}
In this forecast study, we use three dust polarization templates with different complexities. 
\begin{itemize}
\item  \GNILCn: In this model, we first generate  the dust intensity maps $I^{\GNILC}_{\nu}$ \citep{planck-XLVIII:2016} at \ECHO\ frequencies assuming a single component MBB spectrum, 
\begin{equation}
    I^{\GNILC}_{\nu} = \tau_{353}\left (\frac{\nu}{353 }\right )^{\beta_d} B_{\nu} (T_{d}) \, 
\end{equation}
where $\tau_{353}$ is the dust optical depth normalized at 353 GHz, $\beta_d$ is the dust spectral index, $T_d$ is the dust temperature, and $B_{\nu}$ is the planck function. The full-sky maps of $\tau_{353}$, $\beta_d$ and $T_d$ are obtained by fitting the MBB spectrum to the CIB-subtracted \planck\ 353, 545, and 857 GHz intensity maps and IRIS $100\,\mu m$ map. The CIB anisotropies are disentangled from the total intensity maps using the Generalized Needlet Internal Linear Combination (\GNILC) pipeline \citep{Remazeilles:2011}.  

These intensities are converted into the Stokes $Q$ and $U$ parameters as follows,
\begin{align}
& Q^{d}_{\nu} = u_{\nu} f_d \, g_d \, I^{\GNILC}_{\nu} \, \cos\left(2\gamma_{d}\right)\nonumber\\
& U^{d}_{\nu} = u_{\nu} f_d \, g_d \, I^{\GNILC}_{\nu} \, \sin\left(2\gamma_{d}\right),
\label{eq:3.2.1}
\end{align}
where $\gamma_{d}$ is the dust polarization angle arises due to the Galactic magnetic field configuration \citep{Miville-Desch:2008}, $f_d$ is the dust polarization fraction and $g_d$ is the depolarization factor. The polarization angle map ($\gamma_d$) and $g_d$ at scales greater than $20^{\circ}$ are derived from the large-scale model of the Galactic magnetic field \citep{Miville-Desch:2008, ODea:2011}. It is then combined with synchrotron polarization angle ($\gamma_s$) and depolarization factor ($g_s$) maps estimated from the 23 GHz WMAP polarization data and 408 MHz \citep{Haslam:1982} templates to add the structures at intermediate scales between $3^{\circ}$ and $20^{\circ}$. The small-scale ($< 3^{\circ}$) features of $\gamma_d$ and $g_d$ are added using the method described in \citet{Miville:2007}. The combination of different scales to produce high resolution map of $\gamma_d$ is discussed in \citet{PSM:2011}.
We set $f_{d} = 0.15$, which gets reduced to observed polarization fraction $f_d\,g_d$ = 0.05 on average over the sky. The unit conversion factor ($u_{\nu}$) takes into account the conversion from MJy/sr to $\mu K_{\rm CMB}$ units (or thermodynamic temperature units). 

\item \TDdust: \cite{T_Ghosh:2017} and \cite{Adak:2019} build a dust polarization model at 353 \GHz\ at high Galactic latitude using three phases of \hi\ clouds, cold, warm and unstable neutral medium of ISM and phenomenological magnetic field \citep{planck-XLIV:2016}. This model can describe the dust properties within the scale, $40<\ell<160$. We scale those templates at other frequencies with a MBB spectrum using spatially varying $T_d$ of 19.4 $\pm$ 1.4 K and $\beta_d$ with a value of 1.53 $\pm$ 0.02 \citep{planck-XI:2018,planck-XI:2013}. The $\beta_d$ and $T_d$ maps for each \hi\ phase are generated from a Gaussian distribution mean and standard deviation stated above. We do not introduce any correlation between $\beta_d$ and $T_d$ maps for each \hi\ phase across the sky. 
We treat the three \hi\ phases as emitting layers with the same spectral properties/different spectral properties to generate polarized dust maps without/with frequency decorrelation\footnote{Since intrinsic local polarization fraction, $p(r)$, optical depth, $\tau$(r), $T_{d}$, $\beta_{d}$, $\gamma_{d}$, all these parameters change along line-of-sight (LOS), the polarization angle and polarization fraction in each emitting layers are different and function of frequencies. This rotation of the polarization angle along LOS will automatically introduce the depolarization effect \citep{Pelgrims:2021}. However, use of different spectral maps at different layers in this case does not guarantee that the amount of decorrelation will be the same as observed in \cite{planck-XI:2018}.}. We introduce 0.2 \% decorrelation at $\ell = 80$ between 217 and 353 GHz over 24 \% of the unmasked sky. 
We choose this model in order to specifically assess the performance of foreground subtraction with the \COMMANDER\ pipeline in the presence of dust decorrelation across the frequencies.

\item \GINES: We make use of the multi-layer dust model of \cite{gines:2018} based on the dust extinction maps from \cite{Green:2015}, and generate model Stokes $Q$, $U$ parameters with different emission laws in six different layers in their modeling framework. By construction, this model generates decorrelation between the frequency channels. The amount of decorrelation present in the model is 0.5 \% at $\ell = 80$ between 217 and 353 GHz over 70 \% of the unmasked sky. 
\end{itemize}

\begin{figure*}
    \includegraphics[width=\linewidth]{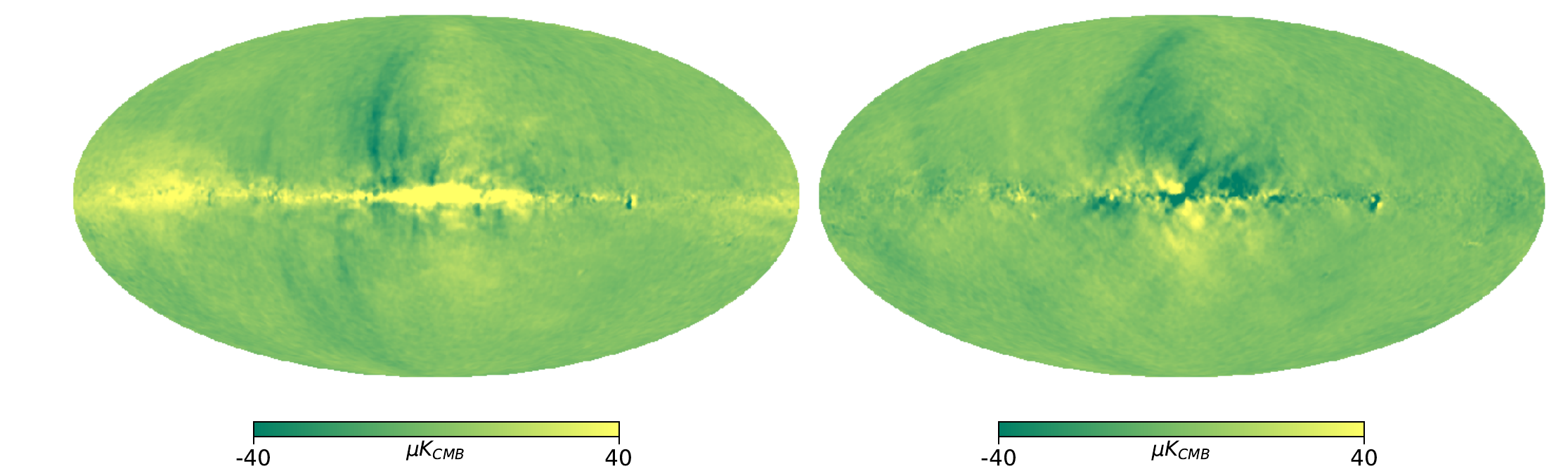} \par
    \includegraphics[width=\linewidth]{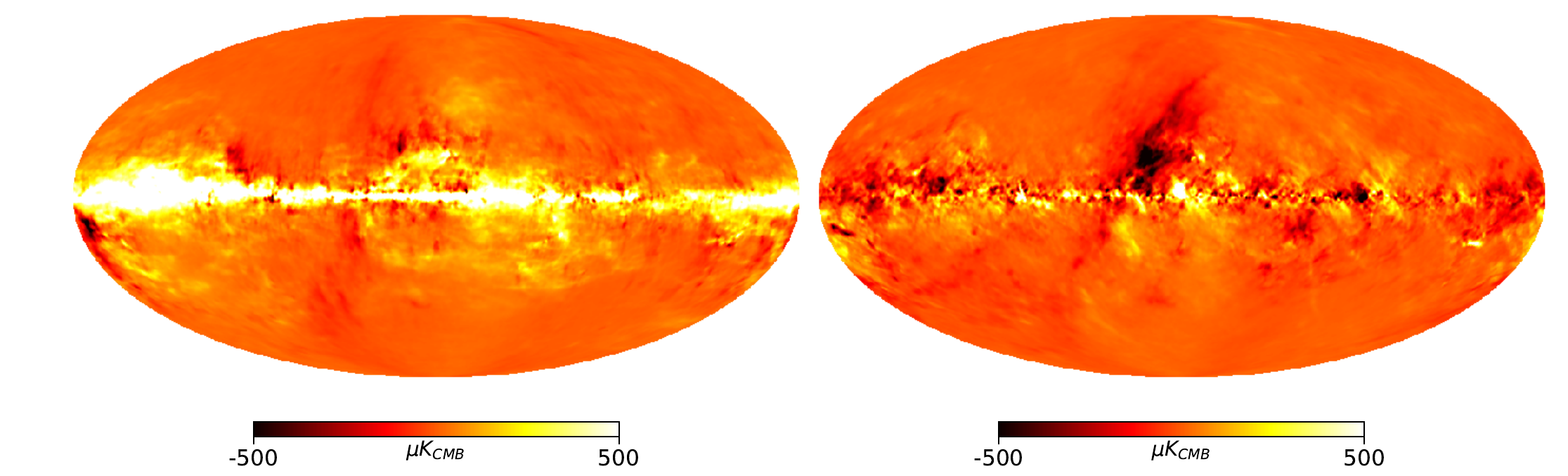} \par

\caption{Simulated Stokes $Q$ and $U$ maps for SET1a at  synchrotron dominated 35\GHz\ channel (\textit{upper panel}) and thermal dust dominated 340\GHz\ channel (\textit{lower panel}). }
\label{fig:B_MODE_MAPS}
\end{figure*}

\subsection{Synchrotron model}
\label{sec:sync}
We consider following three different emission law models of galactic synchrotron emission. For all the models, we use the synchrotron $Q$, $U$ maps obtained from Spectral Matching Independent Component Analysis (\smica) at the reference frequency $\nu_0$ = 30 \GHz\ as the template maps \citep{planck-IV:2018}. 

\begin{itemize}

\item \powerlaw: In this model, the synchrotron $Q^s_{\nu}$ and $U^s_{\nu}$ maps at any frequency $\nu$ are generated by scaling the synchrotron template Stokes $Q_{\nu_{0}}^s$ and $U_{\nu_{0}}^s$ maps (expressed in MJy/sr units) at the reference frequency $\nu_0$ using a power-law model,
\begin{align}
    &Q^{s}_{\nu} \, =\, u_{\nu}Q^s_{\nu_0} \left (\frac{\nu}{\nu_0 } \right )^{\beta_{s}+2} \nonumber\\
    &U^{s}_{\nu} \, =\, u_{\nu}U^s_{\nu_0} \left (\frac{\nu}{\nu_0} \right )^{\beta_{s}+2}, 
\end{align}
where $\beta_s$ is the spatially varying synchrotron spectral index map. It is derived by fitting 408 MHz and WMAP 23 GHz polarization data using model 4 of \citet{Miville-Desch:2008}. The average value of $\beta_s$ is $-3.0$ over the whole sky.
The $u_{\nu}$ denotes the unit conversion factor from MJy/sr to $\mu K_{\rm CMB}$ units. 

\item \curvedpowerlaw: In this model, the synchrotron template maps are scaled to other frequencies using the spectral index,
\begin{equation}
    \beta_{s} = -3.11 + C\,\log \left ( \frac{\nu}{23} \right ).
\end{equation}
We adopt $C = - 0.3$ for frequencies larger than 23 \GHz, which accounts for the steepening of the synchrotron spectrum \citep{Kogut:2007}.

\item \galprop: In this model, the synchrotron template maps are translated to other frequencies using a fixed \galprop\ scaling parameter $\alpha=0.26$ \citep{planck-x:2016}. The synchrotron emission spectrum for $\alpha=0.26$ is simulated from  \galprop 
\footnote{https://galprop.stanford.edu/}\textsuperscript{,}\footnote{https://sourceforge.net/projects/galprop/} code of \cite{Orlando:2013}.

\end{itemize}

\subsection{AME model}
\label{sec:ame}
We model the AME $Q$ and $U$ maps as 
\begin{align}
    & Q^{AME}_{\nu} = p_{AME} I_{\nu}^{AME} \cos(2\gamma_{d})\nonumber\\
    & U^{AME}_{\nu} = p_{AME} I_{\nu}^{AME} \sin(2\gamma_{d}),
\end{align}
where the AME intensity, $I_{\nu}^{AME}$ at the reference frequency 23 \GHz\ is derived from \GNILC\ thermal dust intensity at 353 \GHz\ rescaled by a factor of 0.91\,K/K \citep{planck-xxv:2015}. $I_{\nu}^{AME}$ follows a cold neutral medium model emission law as derived in \cite{Ali:2009} that has been used to extrapolate 23 \GHz\ AME template to \ECHO\ frequencies. We assign to AME emission a constant polarization fraction of $p_{AME} = 1\,\%$, compatible with observational upper limits \citep{Dickinson:2011,Santos:2015}. We assume polarization angle $\gamma_{d}$ to be same as those for thermal dust since both components are correlated.

\subsection{Point source model}
\label{sec:ps}
We consider faint radio and infrared extragalactic sources. The radio sources are taken from radio surveys of PMN/GB6, NVSS, SUMSS at 4.85, 1.4, and 0.843 \GHz\ \citep{PSM:2011}. Then the fluxes of the sources are extrapolated to \ECHO\ frequencies using power-law, $S_{\nu} \propto \nu^{-\alpha}$. Depending on the spectral index, the sources are divided into two categories (the boundary value being $\alpha$ = 0.5); steep- and flat-spectrum class. The polarization fraction is randomly assigned with mean values of 2.7\% and 4.8\% for flat- and steep classes, respectively, consistent with the results of \cite{Ricci:2004}. The faint sources are separated from strong sources following PCCS2 detection limits at 30, 70, 353, 857 \GHz\ giving the flux limit of 0.427, 0.501, 0.304, 0.791 Jy \citep{planck-XXVI:2016}. 

We consider the Infrared Astronomical Satellite (IRAS) faint point sources catalogue \citep{Moshir:1992} and extrapolated them to \ECHO\ frequencies with modified blackbody spectra, $\nu^{b} B(\nu, T)$ with $b = 1.3$ and $T = 35$\,K \citep{Dunne:2000, PSM:2011}. The polarization fraction is assigned randomly drawn from the $\chi^2$ distribution of one degree of freedom with a mean polarization fraction of 1.5 \%, and the polarization angle is drawn from a uniform distribution. 

\begin{table*}
\caption{Summary of the sky components and their parametric model used in simulations.}
\label{table3}
   \begin{centering}
   \begin{tabular}{ p{2.1cm}  p{4.8cm}   p{2.5cm}  p{6.0cm} }
   \hline
   Component & \makecell{Emission law} & Nomenclature & Additional information/Templates\\
   \hline
   \hline
   CMB & \makecell{Blackbody with scaling,\\ $a_{\nu} = \frac{d B_{\nu} (T)}{d T}\vert_{T_{CMB}}$;\\$T_{CMB} = 2.725 $K }&\hspace{0.5cm}& \makecell{$r$ = 0} \\ 
   &&&\\
   Thermal dust & \makecell{MBB } &\GNILCn & \makecell{\planck\ \GNILC\ maps at 353 \GHz\ from \\\cite{planck-XLVIII:2016}} \\
   &&&\\
   & & \TDdust & \makecell{ \hi\ based dust polarization model at \\high galactic latitude  developed in \\ \cite{T_Ghosh:2017} and\\ \cite{Adak:2019} at 353 \GHz}\\
   &&&\\
   & & \GINES & \makecell{ Multi-layer dust model based on dust extinction\\ maps developed in \cite{gines:2018}} \\
   &&&\\
   
   Synchrotron & \makecell{Power-law, spatially varying\\ spectral index with $\langle \beta_{s}\rangle = -3$} & \powerlaw & \smica\ $Q$, $U$ maps from \cite{planck-IV:2018} at 30 \GHz \\
   &&&\\
   & \makecell{Frequency dependent spectral\\ index; $\beta_{s} = -3.11 + C\, \log(\frac{\nu}{23})$ \\ with curvature, $C = -0.3$\\ at 23\GHz}& \curvedpowerlaw & \smica\ $Q$, $U$ maps from \cite{planck-IV:2018} at 30 \GHz \\
   &&&\\
   & \makecell{ \galprop\ scaling; $(\frac{\nu}{30})^{2}\frac{f_{s}(\frac{\nu}{\alpha})}{f_{s}(\frac{30}{\alpha})}$ \\ with constant $\alpha$ = 0.26 and $f_{s}(\nu)$ is\\ taken from external template\\ generated from \galprop\ code} & \galprop & \smica\ $Q$, $U$ maps from \cite{planck-IV:2018} at 30 \GHz\\
   &&&\\
   Spinning dust &\makecell{ CNM emission law with 1 \% polarization\\ fraction and dust polarization angle}& &\makecell{ \planck\ thermal dust intensity at 353\GHz\\ \citep{planck-XLVIII:2016}  scaled\\ at 23 \GHz\ with correlation coefficient of 0.91 K/K} \\
   &&&\\
   Point-sources &\makecell{Sources from radio surveys\\ extrapolated with power laws;\vspace{0.5cm} \\ 
   IRAS survey modelled with\\ modified blackbody emission laws.}&\hspace{1cm}&\makecell{ Radio sources have median polarization fraction of\\ 2.7 \% and 4.8 \% for two class of power-laws; \vspace{0.5cm} \\IR sources are taken from IRIS data and \\having mean polarization fraction of 1.5 \%} \\
   &&&\\
   \hline\\
   \end{tabular}
   \end{centering}
\end{table*}
%

\subsection{Sky simulations}\label{sec:psm}

We use the recent version of the Planck Sky Model\footnote{http://www.apc.univ-paris7.fr/~delabrou/PSM/psm.html} (PSM v2.0.2, \citealt{PSM:2011}) to simulate full-sky maps for most of the sky configurations. In addition, we use the polarized dust templates of Stokes $Q$ and $U$ maps from \GINES\ \citep{gines:2018} and \TDdust\ \citep{T_Ghosh:2017, Adak:2019} models as described in Section~\ref{sec:dust}. Table~\ref{table3} summarizes all the sky components and their models included in the simulations. To test the efficiency of the two component separation methods in the recovery of $r$,  we choose to work with low \healpix\ pixel resolution of \Nside\ = 512 for \NILC\ and \Nside = 256 for \COMMANDER. We first simulate all the component maps (CMB, foreground and instrumental noise) at \Nside\ = 512. For consistency with our choice of \Nside, CMB and foreground maps at \Nside\ = 512 are smoothed at the respective beam resolution specified in Table~\ref{table2} for the frequency channels below 65 \GHz\ and a FWHM = 20\parcm\ for rest of the frequency channels. These maps are used for \NILC\ analysis. For \COMMANDER, we apply additional smoothing to make all the maps at a common beam resolution of 60\parcm\ FWHM and downgraded to \Nside\ = 256.

In Table~\ref{table4}, we summarize the set of sky configurations together with the different choices of foreground cleaning algorithms employed. We include the polarized thermal dust and synchrotron emissions to all the set of simulations  that correspond to a different models of dust and synchrotron emission. 
 For SET3a-b sky configurations, we choose to work with \Nside = 128 maps with 60\parcm\ FWHM beam smoothing. Since the \TDdust\ templates are produced at $\Nside = 128$, the other foreground components are further downgraded to \Nside\ = 128 to match the \healpix\ pixelization scheme. 

We generate the two half-mission (HM1 \& HM2) and one full-mission Gaussian white noise realizations at each frequency.  For simulating the half-mission noise, we use the $\sqrt{2}$ times the r.m.s noise level listed in Table~\ref{table2}. Noise maps are smoothed to final beam FWHM for each channel. Finally, we coadd CMB, foreground and noise realization maps at each frequency to produce the two half-mission data sets of the observed sky at respective \ECHO\ frequencies. The final sky maps are expressed in thermodynamic unit. Figure~\ref{fig:B_MODE_MAPS} presents the full-sky Stokes $Q$ and $U$ maps at 35 \GHz\ (\textit{upper panel}; synchrotron dominated channel) and 340 \GHz\ (\textit{lower panel}; thermal dust dominated channel) for the sky configuration SET1a.

\begin{table*}
\caption{Set of simulations and pipelines used in the analysis. The tick and cross symbols indicate which components are added and excluded respectively for different configurations. The dust and synchrotron models used can be identified using nomenclatures listed in Table~\ref{table3}. }
\label{table4}
   \begin{centering}
   \begin{tabular}{ p{1.cm}  p{1.2cm} p{.9cm}  p{2.5cm}  p{2.5cm} p{1.2cm} p{1.2cm} p{1.2cm} p{1.5cm}}
   \hline
    Sim.ID&\multicolumn{2}{c}{Pipeline}\hspace{0.5cm} & \hspace{0.5cm}Dust & Synchrotron& AME & point-sources & delensing \hspace{0.3cm} & Decorrelation \hspace{0.2cm}\\\cline{2-3}
    &\COMMANDER & \hspace{0.2cm}\NILC &&&&&\\
    \hline
    \hline
    SET1a&\hspace{0.2cm}\cmark&\hspace{0.2cm}\cmark &\GNILCn& \galprop& \xmark&\xmark&\xmark &\xmark\\
    SET1b&\hspace{0.2cm}\cmark&\hspace{0.2cm}\cmark &\GNILCn& \galprop& \cmark&\xmark&\xmark&\xmark\\
    SET1c&\hspace{0.2cm}\cmark&\hspace{0.2cm}\cmark&\GNILCn& \galprop& \cmark&\cmark&\xmark&\xmark\\
    SET1d&\hspace{0.2cm}\cmark&\hspace{0.2cm}\cmark &\GNILCn& \galprop& \xmark&\xmark&\cmark &\xmark\\
    &&&&&&&\\
    SET1e&\hspace{0.2cm}\cmark&\hspace{0.2cm}\cmark &\GNILCn& \powerlaw & \cmark&\cmark&\xmark&\xmark \\
    SET1f&\hspace{0.2cm}\cmark&\hspace{0.2cm}\cmark&\GNILCn& \curvedpowerlaw & \cmark&\cmark&\xmark&\xmark \\
    &&&&&&&\\
    SET2a&\hspace{0.2cm}\cmark&\hspace{0.2cm}\cmark &\GINES& \galprop & \cmark&\cmark&\xmark&\cmark \\
    SET2b&\hspace{0.2cm}\cmark&\hspace{0.2cm}\cmark &\GINES& \powerlaw & \cmark&\cmark&\xmark&\cmark \\
    SET2c&\hspace{0.2cm}\cmark&\hspace{0.2cm}\cmark &\GINES& \curvedpowerlaw & \cmark&\cmark&\xmark&\cmark\\
    
    &&&&&&&\\
    SET3a&\hspace{0.2cm}\cmark& \hspace{0.2cm}\xmark&\TDdust   & \galprop & \cmark&\cmark&\xmark&\xmark \\ 

    SET3b&\hspace{0.2cm}\cmark&\hspace{0.2cm}\xmark&\TDdust & \galprop & \cmark&\cmark&\xmark&\cmark  \\ 
    
    \hline
\end{tabular}
\end{centering}

\end{table*}
%


\section{Methods}
\label{sec:method}
\subsection{Component separation pipelines}
\label{sec:pipelines}
In this section, we briefly describe the two component separation methods (\COMMANDER\ and \NILC) that have been implemented to recover the $B$-mode of CMB from multi-frequency simulations of the sky. As far as the implementation of these methods is concerned, these two methods differ a lot from each other, but, they are complementary to each other. While \NILC\ exploits the black body frequency scaling of the CMB and minimizes the variance of the linear mixture of the sky operating on spherical wavelet domain with minimum assumption of the foreground emissions, the \COMMANDER\ makes use of the prior knowledge of the foreground emissions characterized by the parameters which are fitted using Bayesian techniques operating on pixel domain.

\subsubsection{\COMMANDER\ Implementation}
\COMMANDER\ assumes that the observed sky can be described using parametric model and is a linear superposition of the foreground emissions and instrumental noise superimposed on true CMB. The central idea behind this method is the use of Gibbs sampling technique \citep{Wandelt:2004} to sample from 
joint posterior distribution by cycling through conditional distributions \citep{Eriksen:2004, Eriksen:2008}.
We use the publicly available version of \COMMANDER\ code, called \COMMANDER1\footnote{\url{https://github.com/Cosmoglobe/Commander}}. We fit the HM1 and HM2 simulations at each pixel $p$ with the model consisting of CMB, dust and synchrotron emissions. The foregrounds (dust and synchrotron) spectra are modelled in terms of few free parameters characterizing their amplitudes and spectral indices,
\begin{align}
    d_\nu(p) &= s^{cmb} (p) + \sum_{i=1}^{N_{foreground}} F^{i}_{\nu}(\beta_i (p)) f^i(p) + n_{\nu}(p),
\end{align}
where $s^{cmb}$ and $f^i$ are the CMB and foreground amplitudes respectively at some reference frequency. In thermodynamic units, the CMB amplitude is constant across the frequencies. We choose 28 \GHz\ and 340 \GHz\ as the reference frequency for synchrotron and dust respectively. The frequency scaling coefficients $F_{\nu}^i(\beta_i (p))$ are set by the parametric model of the spectral emission laws of the foregrounds, and $n_{\nu}$ is the instrumental noise. We fit the MBB spectrum of the dust and different emission law models (power-law, curved power-law, and \galprop) for the synchrotron to each of the half-mission data sets. We use Gaussian priors for dust temperature, $T_{d} = 19 \pm 0.3$\,K, dust spectral index, $\beta_{d} = 1.6 \pm 0.3$ and synchrotron spectral index, $\beta_s = -3.0 \pm 0.2$ along with Jeffreys priors \citep{Jeffreys:1939}.

The detailed sampling method has been described in \cite{Eriksen:2008}. The method is computationally demanding, and hence can work only with the low-resolution \healpix\ maps at \Nside\ = 256 and 128. We draw the samples of the parameters over the full-sky for all the sky configurations, except SET3a and SET3b. 

For SET1a-d, the data model we fit within the \COMMANDER\ framework is the same as the one used for the sky simulations. However, for SET2a-c and SET3a-b, there is a mismatch between the simulated maps and the data model due to the presence of the dust decorrelation. This model mismatch could result into a excess bias on the estimation of $r$. In principle, we could reduce this bias by fitting the dust decorrrelation parameters in the \COMMANDER\ framework. As our polarized dust simulations capture the sky complexity, it is not simple to find a parametric model that captures the dust decorrelation fully. 

\subsubsection{\NILC\ Implementation}
\label{sec:NILC Implementation}
On contrary to \COMMANDER\ method, \NILC\ is a blind component separation method. It is implementation of ILC of the frequency channels with minimum error variance, on a frame of spherical wavelets called needlets, allowing localized filtering in both pixel domain and harmonic domain.

The \NILC\ pipeline \citep{J.Delabrouille:2009,Basak_and_Delabrouille:2012,Basak:2013} is designed to be applicable to scalar fields on sphere. Therefore, we work separately on $E$ and $B$ maps decomposed from $Q$ and $U$ maps. Prior to the implememtaion of the \NILC\ to the simulated maps, polarization maps of the sky at each frequency are de-convolved to a common beam resolution which is Gaussian beam of FWHM = $20^{\prime}$ in our case. The common resolution sky maps are then decomposed into a set of needlet coefficients for different needlet scales. The needlet coefficients at different needlet scale $j$ are obtained using needlet filters $h^{j}_{l}$ designed as follows:

\begin{equation}
    \centering
    h^j_{\ell} = \left\{ \begin{tabular}{cl}
    $\cos\left[ \left(\frac{\ell^{j}_{peak}- \ell}{\ell^{j}_{peak}- \ell_{min}^{j}}\right)\frac{\pi}{2}\right]$ & \text{for } $\ell_{min}^{j}$ $\leq$ $\ell$ $< \ell_{peak}^{j}$ \\
    &\\
    $1$ & \text{for } $\ell =$ $\ell_{peak}^{j}$\\
    &\\
    $\cos\left[ \left(\frac{\ell-\ell^{j}_{peak} }{\ell^{j}_{max}- \ell_{peak}^{j}}\right)\frac{\pi}{2}\right]$ & \text{for } $\ell_{peak}^{j}<$ $\ell$ $\leq$ $\ell_{max}^{j}$. 
    \end{tabular} \right.
\end{equation}
\begin{table}
\caption{List of needlet bands used in the analysis.}  \centering \begin{tabular}{c c c c c} \hline\hline
  Band index & $l_{min}$ & $l_{peak}$ & $l_{max}$ & $N_{side}$ \\ [5ex]
  \hline 1 & 0 & 0 & 50 & 32\\ 2 & 0 & 50 & 100 & 64 \\ 3 & 50 & 100 &
  200 & 128 \\ 4 & 100 & 200 & 300 & 128 \\ 5 & 200 & 300 & 400 & 256
  \\ 6 & 300 & 400 & 500 & 512 \\ 7 & 400 & 500 & 600 & 512 \\ 8 & 500 
  & 600 & 700 & 512 \\ 9 & 600 & 700 & 800 & 512 \\ 10 & 700 & 800 & 900 & 512 \\ 11 & 800 & 900 & 1000 & 512 \\ [1ex] \hline
\end{tabular} 
\label{tab:needlet-bands} 
\end{table}
\begin{figure}
    \includegraphics[width=8.4cm]{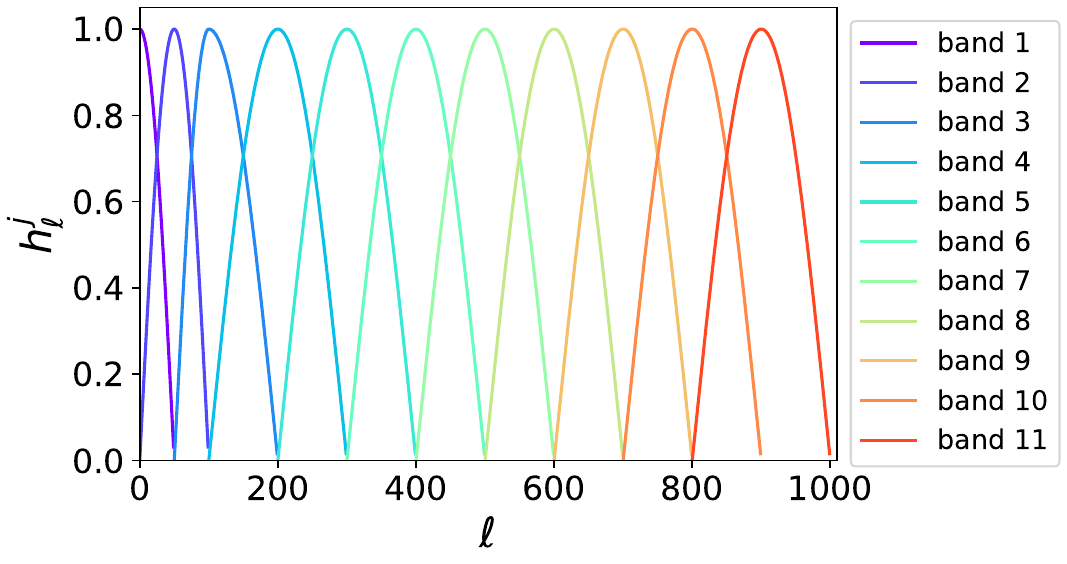}      
    \caption{The needlet filter function $h^j_l$ for eleven needlet bands.}
    \label{fig2:NILC_bands}
\end{figure}

For each scale $j$, the filter has a compact support between ranging between the multipoles $l^{j}_{min}$ and $l^{j}_{max}$ with a peak at $l^{j}_{peak}$. For our purpose, we have used $11$ needlet bands listed in Table~\ref{tab:needlet-bands} and corresponding filters are shown in Fig.~\ref{fig2:NILC_bands}. The needlet coefficients are computed at \healpix\ grid points with a resolution, \Nside\ equal to the smallest power of 2 larger than $l^{j}_{max}/2$. The corresponding \Nside\ for different needlet scales are listed in Table~\ref{tab:needlet-bands}.

 For each scale, the \NILC\ weights are determined by minimizing the variance of the linear combination of the sky in such a way that the foreground and noise component present in sky is minimized while the CMB component remain unaffected. Finally, the CMB is extracted by linearly combining the sky at each frequency through the \NILC\ weights. Thus, the contribution of each frequency channel towards the reconstructed CMB map can be easily demonstrated through the \NILC\ weights.

\subsection{Galactic masks}
\label{sec:mask}
\begin{figure*}
    \begin{multicols}{3}
       \includegraphics[width=5.9cm]{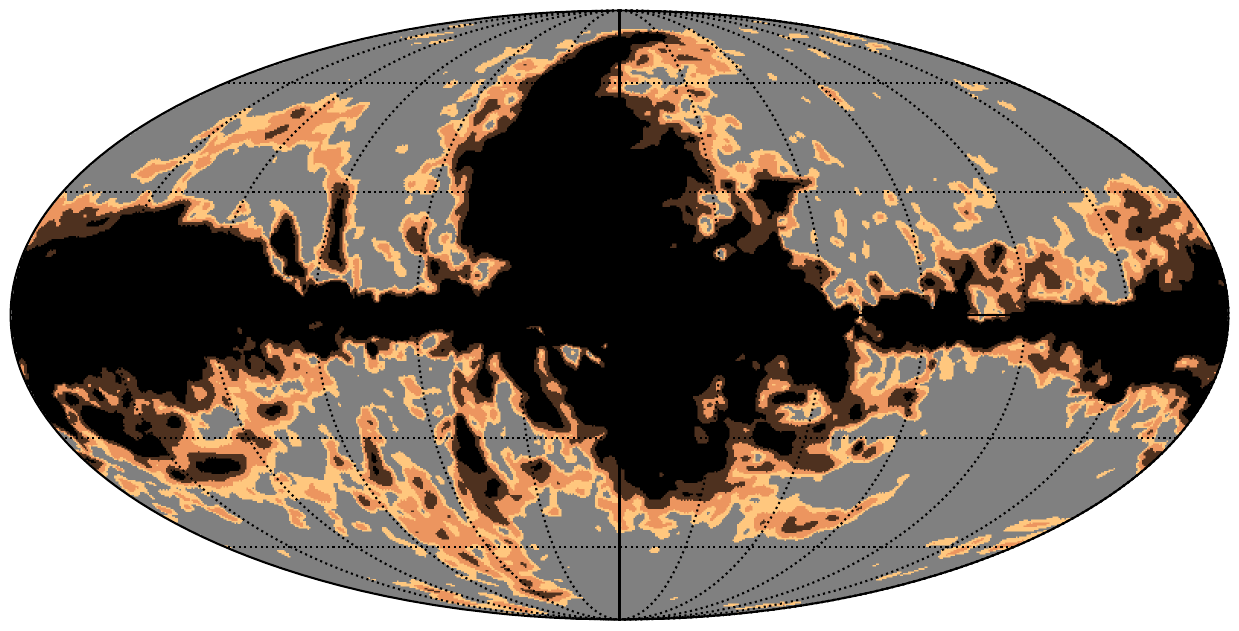}\par
       \includegraphics[width=5.9cm]{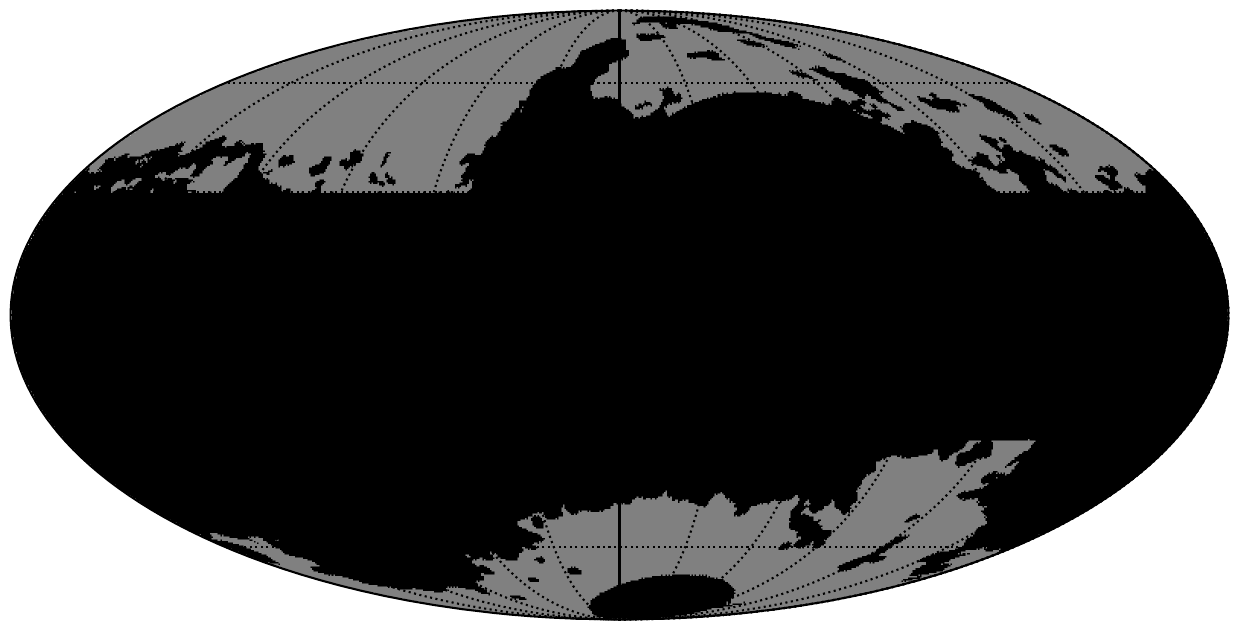}\par
       \includegraphics[width=5.9cm]{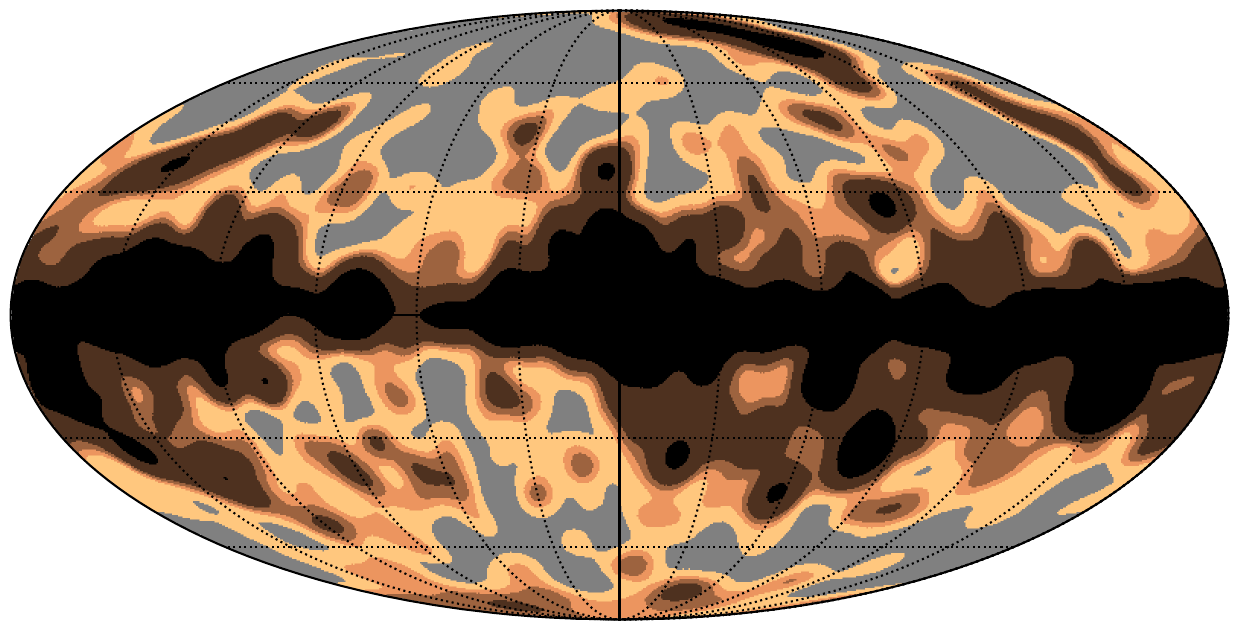}
    \end{multicols}
    \caption{Left panel: Series of Galactic mask used analysis of CMB map recovered from \COMMANDER\ for SET1a. The region shown in \textit{darkest brown} is masked out to retain 70 \% sky. The regions shown in increasingly fainter \textit{brown} colors are incrementally masked out regions, corresponding in turn to retain 60 \%, 50 \%, 40 \% and 30 \% sky fractions respectively. Middle panel: Galactic mask comprising 24 \% sky fraction at \Nside\ = 128 that is used in analysis of recovered CMB maps from \COMMANDER\ for SET3a-b simulations. The region shown in \textit{darkest brown} color is masked out. Right panel: \NILC\ Galactic masks for $B$-mode for SET1a simulation. The \textit{ darkest brown} region masks 20 \% of the sky. The regions with increasingly fainter \textit{brown} colors mask out 40 \%, 50 \%, 60 \% and 80 \% of the sky respectively.}
    \label{fig:commander_nilc_mask}
\end{figure*}

To minimize the impact of foreground residuals on the measurement of $r$, we apply a conservative mask. We put thresholds on the synchrotron and dust polarization intensity maps to prepare the confidence mask for \COMMANDER\ \citep{Bennett:2013, planck-x:2016, CORE-B:2018}. 
To generate the \COMMANDER\ confidence mask, we use 5\deg\ FWHM beam smoothed synchrotron Stokes $Q^s_{\nu_0}$ and $U^s_{\nu_0}$ template maps at reference frequency $\nu_0=30$ \GHz\ and extrapolate to 75 \GHz\ using a power-law spectrum with $\beta_{s}  = -3.0$. Next we compute the polarization intensity, $P^{s}=\sqrt{(Q_{75}^s)^2 +(U_{75}^s)^2}$, from the extrapolated Stokes $Q$ and $U$ maps and mask pixels where $P^s$ is greater than 2 - 6 times of 0.16 \muK. The reference cutoff of 0.16 \muK\ is the r.m.s fluctuations of the CMB polarization intensity map smoothed at 5\deg\ FWHM. The dust polarization intensity thresholds is determined in the similar fashion by extrapolating 353 \GHz\ \GNILC\ dust map to 75 \GHz\ using the MBB spectrum with a fixed $T_{d} = 19.4$\,K and $\beta_{d} = 1.6$. We use the same 
to define the dust masks. The final mask is constructed from the union of the dust and synchrotron masks. The left panel of Fig.~\ref{fig:commander_nilc_mask}, shows the set of Galactic mask used for SET1a that retains 30 - 70 \% sky fraction. The regions shown in brown colors of different shades are masked out to retain different sky fraction.

Similarly, we prepare the Galactic mask for the sky configuration SET3a-b at  \Nside\ = 128 by choosing the threshold of 0.48 \muK, which is three times the r.m.s fluctuations of the CMB polarization intensity map smoothed at 5\deg\ FWHM. The analysis mask is displayed in the middle panel of Fig.~\ref{fig:commander_nilc_mask} that retains 24 \% of the sky.

The \NILC\ confidence masks are constructed by examining the foreground residuals present in the recovered CMB $B$-mode map. We prepare the foreground residual $B$-mode map by propagating the \NILC\ weights obtained from the analysis of the total frequency maps to foreground maps only at \ECHO\ frequencies. We square the foreground residual map, smoothed it to 9\deg\ FWHM beam resolution, and choose appropriate threshold values to retain 20 \% to 80 \% of the total sky fraction. 
The set of \NILC\ masks used for the analysis of sky configuration SET1a are displayed in the right panel of Fig.~\ref{fig:commander_nilc_mask}. The different brown shaded regions are masked out for retaining different sky fractions. 

\subsection{Power spectrum estimation and Likelihood Analysis}
\label{sec:likelihood}
We compute the $BB$ power spectra cross-correlating the recovered $B$-mode CMB maps for HM1 and HM2 over the masked sky and then apply the power spectrum based likelihood approach to estimate $r$.
 We apply pseudo-$C_{\ell}$ estimator \citep{Hivon:2002} from \namaster\ code\footnote{\url{https://github.com/damonge/NaMaster}} \citep{Namaster:2019}. We use apodized mask fallowing Gaussian apodization scheme with FWHM = 1\deg. \NILC\ returns directly the foreground cleaned CMB $B$-mode maps. 
 For \COMMANDER, we first decompose the full-sky foreground cleaned CMB $Q$ and $U$ maps to full-sky $E-$ and $B$-mode maps and then use only the $B$-mode maps in order to avoid the leakage from $E$- to $B$-modes over the masked sky. 
We adopt the constant binning scheme of band width $\Delta \ell$ = 9 and ($2\ell$+1) weighted band power in $k^{th}$ band given by
\begin{align}
    \hat{C}_{k}^{BB} = \sum_{\ell_{min}(k)}^{\ell_{max}(k)} \frac{(2\ell + 1)}{\sum_{\ell =  \ell_{min}(k)}^{\ell_{max}(k)} (2\ell + 1)} {C}_{\ell}^{BB},
    \label{eq3:seclikelihood}
\end{align}
where ${C}_{\ell}^{BB}$ is the unbinned cross power spectrum (hereafter, by $C_{\ell}^{BB}$, we will indicate band power). For simulations in SET3a and SET3b, \COMMANDER\ is applied over the masked sky. To avoid the issue of $E$-to-$B$ leakage over the masked sky, we adopt the pure-$B$ estimator implemented in the \namaster\ code which minimizes the leakage with additional sample variance \citep{SMITH:2006}.

Next, we pass the estimated the binned power spectra to the log-likelihood,
\begin{align}
\label{eq:likelihood}
    &-2ln\mathcal{L} = constant \,+ \nonumber \\
    & \sum_{k,k^{'}} \left[ \hat{C}_{k}^{BB} - C_{k}^{BB,model} (r, A_{L})\right ] \Sigma_{k,k^{'}}^{-1}\left[ \hat{C}_{k^{'}}^{BB} - C_{k^{'}}^{BB,model} (r, A_{L})\right ] ,
\end{align}
where $k$, $k^{'}$ indicate the bands and $\Sigma_{k,k^{'}}$ is $BB$ band covariance matrix. 
The band covariance matrix is estimated following similar binning scheme as in Equation~\ref{eq3:seclikelihood}, 
\begin{align}
    \Sigma_{k,k^{'}} = \frac{\sum_{\ell = \ell_{min}(k)}^{\ell_{max}(k)} \sum_{\ell^{'} = \ell^{'}_{min}(k)}^{\ell^{'}_{max}(k)} (2\ell + 1) (2\ell^{'} + 1) \Xi_{\ell \ell^{'}}^{BB,BB}}{\sum_{\ell = \ell_{min}(k)}^{\ell_{max}(k)} \sum_{\ell^{'} = \ell^{'}_{min}(k)}^{\ell^{'}_{max}(k)} (2\ell + 1) (2\ell^{'} + 1)}
     \label{eq4:seclikelihood}
\end{align}
where $\Xi_{\ell \ell^{'}}^{BB,BB} $ \citep{XSPECT:2005} is the $BB$ covariance matrix estimated using \namaster\ code. In order to take into account the impact of residual foreground and noise in covaraince matrix, we consider both auto- and cross-power spectra in covariance matrix estimation.

We assume, at likelihood level, $BB$ power spectra can be expressed as combination of theoretical tensor and lensing modes as,

\begin{align}
    C_{k}^{BB,model} (r, A_{L}) = \frac{r}{0.01} C_{k}^{BB,tensor} (r = 0.01) + A_{L} C_{k}^{BB,lensing},
    \label{eq2:seclikelihood} 
\end{align}
where the reference tensor power spectrum, $C_{k}^{BB,tensor}$ at $r$ = 0.01 and lensing power spectrum, $C_{k}^{BB,lensing}$ are held fixed to baseline $\Lambda$CDM model parameters \citep{planck-IV:2018}. The reference $C_{\ell}$s are binned using same weighting scheme shown in Equation~\ref{eq3:seclikelihood}. We assume that impact of delensing can be compressed to a single scale invariant parameter $A_{L}$. We fix $A_{L}=1$  (for no delensing) and 0.16 (for 84 $\%$ delensing) in Equation~\ref{eq2:seclikelihood} while estimating the maximum probable value $r_{mp}$ and its uncertainty $\sigma(r_{mp})$. By maximising the log-likelihood given in Equation~\ref{eq:likelihood}, we analytically estimate the best-fit value of $r_{mp}$ given by,

\begin{equation}
    r_{mp} = 0.01 \times \frac{\sum_{k, k^{'}} \left[ \hat{C}_{k}^{BB} - A_{L} C_{k}^{BB,lensing}\right ] \Sigma_{k,k^{'}}^{-1} C_{k^{'}}^{BB,tensor} }{ \sum_{k, k^{'}}\left[  \hat{C}_{k}^{BB,tensor} \Sigma_{k,k^{'}}^{-1} C_{k^{'}}^{BB,tensor} \right ] }.
    \label{eq:rmplikelihood}
\end{equation}

We can compute the Fisher matrix using $F_{ij} = - \Bigg \langle \frac{\partial^{2} ln\mathcal{L}}{\partial \theta_{i} \partial \theta_{j}} \bigg \rangle$ ($r$ and $A_{L}$ are denoted by general notation $\theta$) evaluated at  best-fit parameters. The 1$\sigma$ error bar of the parameters are given by $\sqrt{F_{ii}^{-1}}$. Using Equation~\ref{eq:likelihood} in Fisher matrix, we obtain the 1$\sigma$ error bar of $r_{mp}$ as,

\begin{equation}
    \sigma(r_{mp}) = 0.01 \times \sqrt{\left[  \hat{C}_{k}^{BB,tensor} \Sigma_{k,k^{'}}^{-1} C_{k^{'}}^{BB,tensor} \right ]^{-1}}.
    \label{eq:Fmatrixlikelihood}
\end{equation}

\section{Forecast Results}
\label{sec:results}
The recovered CMB maps obtained using \COMMANDER\ and \NILC\ are expected to contain residuals of foreground and instrumental noise. In case of polarization of CMB, the level of the residual can be high enough to largely bias the measurement of angular power spectra depending on foreground complexity and hence the measurement of tensor-to-scalar ratio. Lensing $B$-mode of CMB adds to this complication because it appears as nuisance to primordial $B$-mode of CMB. In this section, we discuss the impact of  foreground residuals on biasing $r$ measurement for various complexity of the foreground models and lensing corrections.

\begin{figure}
    \includegraphics[width=\linewidth]{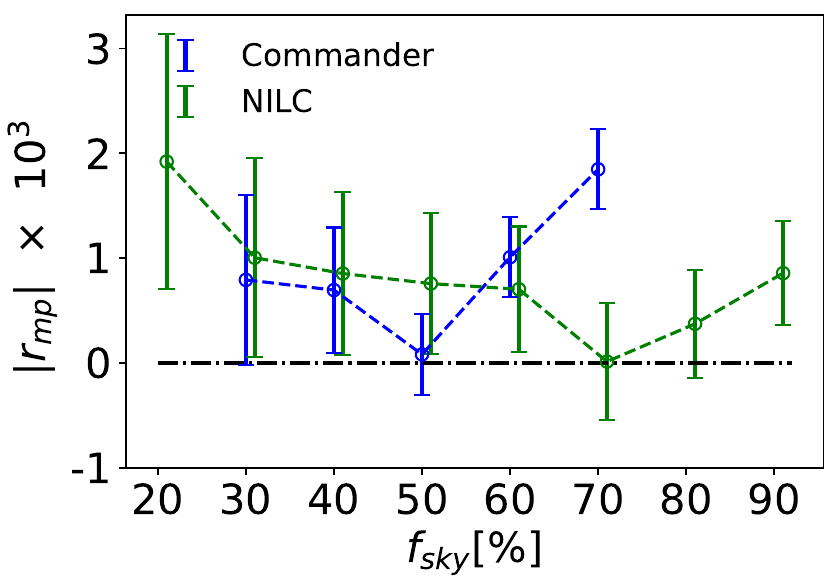} 
\caption{The changes in $r$ bias and uncertainty estimated from foreground cleaned CMB maps by \COMMANDER\ (\textit{blue}) and \NILC\ (\textit{green}) with different sky fractions for simulation in SET1a. Black dashdotted line is input $r$ = 0. The $r$ bias reduces with reducing sky fraction with the cost of small degradation of $r$ sensitivity up to $f_{sky} = $ 50 \% (for \COMMANDER) and 70 \% (for \NILC). Further reduction of sky fraction introduces larger bias  due to loss of information at large angular scales.}
\label{fig:posterior_fsky}
\end{figure}

\begin{table}
  \caption{The maximum probable value of $r$ and its associated uncertainties obtained from the posterior distribution for different choices of $\ell_{max}$ cutoff for SET1a. The sky fraction is fixed at 40 \% for \COMMANDER\ analysis. Uncertainties are consistently improved with addition of power from higher $\ell$ modes up to $\ell_{max} \sim$ 130, since most of the constraining power on $r$ is localized near recombination bump at $\ell \sim$ 100.} 
  \label{tab:lmaxcutoff}
   \begin{centering}
   \begin{tabular}{ p{2.2cm}    p{1cm} p{1.3cm}    p{1cm} p{1.3cm}   }
   \hline
     $\ell_{max}$& \multicolumn{2}{c}{\COMMANDER}& \multicolumn{2}{c}{\NILC}\\ \cline{2-5}
      &$r_{mp}\times 10^3$ & $\sigma(r)\times 10^3$   & $r_{mp}\times 10^3$ & $\sigma(r)\times 10^3$  \\
    \hline
    \hline
             24(near reionization bump)&-1.06&0.82&-0.12&1.69 \\
             42&-0.77&0.42&1.37&1.13\\
             60&-0.51&0.42&0.24&0.89\\
             78&-0.35&0.40&-0.07&0.77\\
             96(near recombination bump)&-0.15&0.39&-0.26&0.72\\
             114&-0.15&0.39&-0.42&0.69\\
             132&-0.14&0.39&-0.54&0.68\\
             150&-0.11&0.39&-0.56&0.68\\
             180&-0.08&0.39&-0.62&0.68\\
             249&-&-&-0.68&0.67\\
             348&-&-&-0.76&0.68\\
             447&-&-&-0.75&0.67\\
             501&-&-&-0.75&0.67\\
             591&-&-&-0.76&0.67\\
            
    \hline
\end{tabular}
\end{centering}
\end{table}
\begin{figure}
  \includegraphics[width=8.4cm]{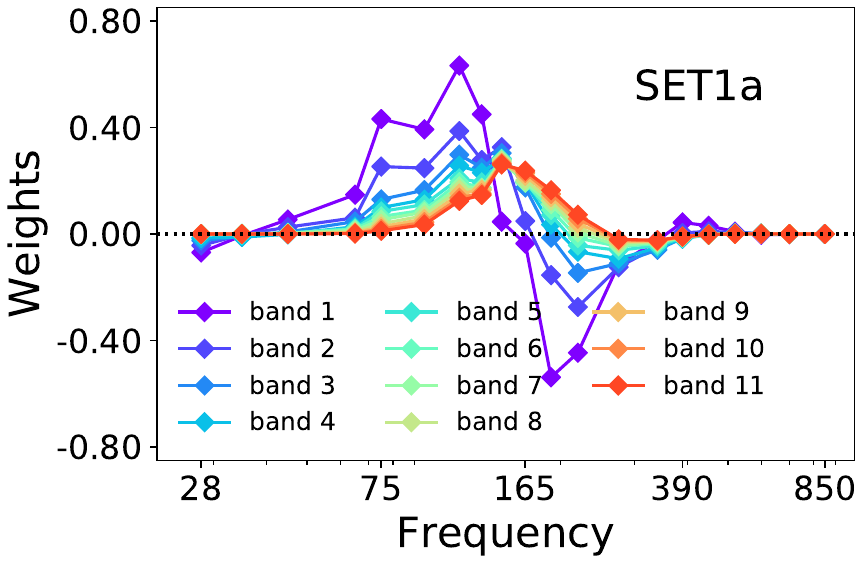}     

    \caption{Distribution of the full sky average values of the \NILC\ weights across different frequency channels. All the 11 \NILC\ bands have been displayed in the figure for the SET1a sky model.}
\label{fig1:NILC_weights_set1a}
\end{figure}

\begin{figure*}
\begin{multicols}{2}
    \includegraphics[width=\linewidth]{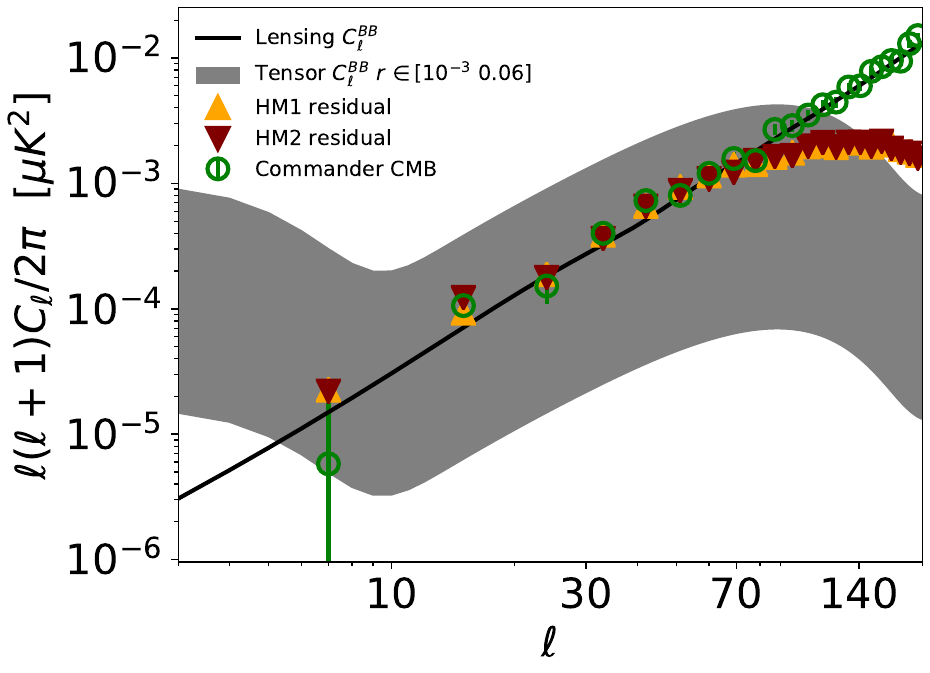} \par
    \includegraphics[width=\linewidth]{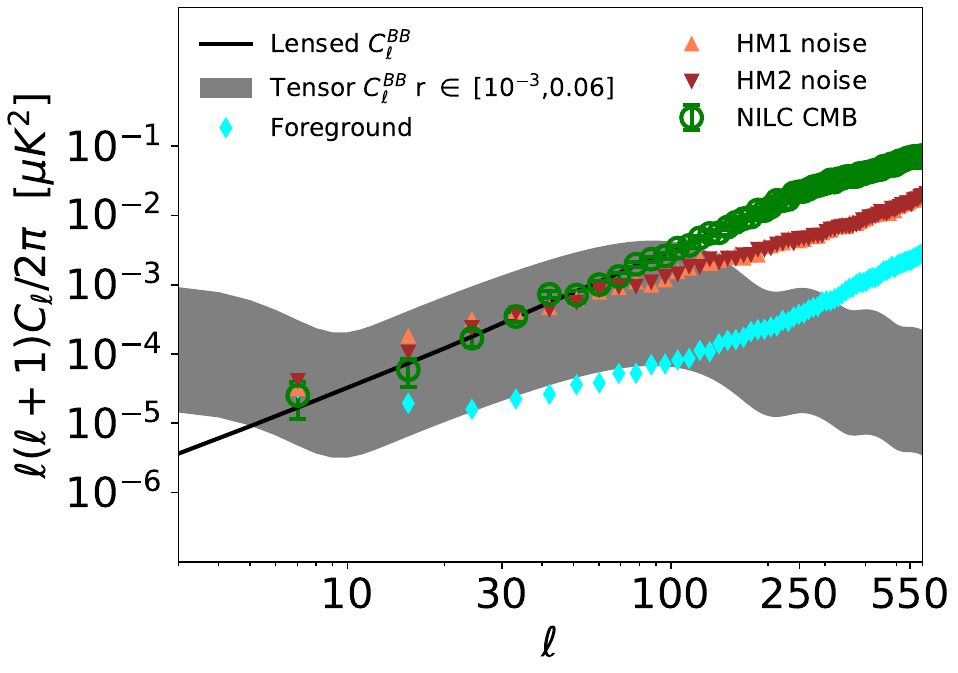} \par
    \end{multicols}
\caption{$BB$ cross-power spectra (\textit{green circles}) estimated over 50\% \& 70\% sky fraction from HM1, HM2 recovered maps using \COMMANDER\ (\textit{left panel}) \&  \NILC\ (\textit{right panel}) respectively for configuration SET1a. The 1$\sigma$ error bars are the quadratic sum of cosmic variance of CMB $C_{\ell}$ and uncertainty introduced by residual noise and foregrounds as estimated by \namaster. \textit{Black solid} line is the theoretical lensed $BB$ power spectrum, grey-shaded region is theoretical tensor $BB$ power spectra for range of $r$ $\in [10^{-3}, 0.06] $. Contribution of foreground + noise residual leakage to two half-mission maps recovered by \COMMANDER\ are shown in \textit{orange} and \textit{maroon} triangles. For \NILC, the residual foreground (\textit{cyan diamonds}) and noise power spectra for two half-missions (HM1 in \textit{maroon} and HM2 in \textit{orange}) are displayed for reference. }
\label{fig1:examplecllikelihood}
\end{figure*}

\begin{table*}
\caption{{Results of maximum likelihood estimation of $r$ from $BB$ power spectrum estimated from foreground-cleaned maps using \NILC\ and \COMMANDER\ pipelines. We list best-fit foreground bias on $r$, 1$\sigma$ uncertainties and $\chi^2$/dof for different foreground setups.}}
\label{table5}
   \begin{centering}
   \begin{tabular}{  p{2.7cm}   p{2.0cm} p{2.0cm}  p{2.0cm}  p{2.0cm} p{2.0cm}  p{2.0cm}}
   \hline
     Sim.ID& \multicolumn{3}{c}{\NILC}& \multicolumn{3}{c}{\COMMANDER}  \\ \cline{2-7}
                  &\hspace{0.5cm} $r_{mp}\times 10^{3}$ &\hspace{.5cm} $\sigma(r_{mp})\times 10^{3}$ &\hspace{0.5cm} $\chi^2$/dof  &\hspace{0.5cm} $r_{mp}\times 10^{3}$ &\hspace{0.5cm} $\sigma(r_{mp})\times 10^{3}$  &\hspace{0.5cm} $\chi^2$/dof  \\
            \hline
            \hline
            &&&&&&\\
             SET1a & \hspace{0.5cm}-0.76 &\hspace{0.5cm} 0.67 &\hspace{0.5cm}0.60 & \hspace{0.3cm} -0.08&\hspace{0.4cm} 0.39& \hspace{0.5cm} 0.95 \\
            SET1b &\hspace{0.4cm} -0.55 &\hspace{0.5cm} 0.68 & \hspace{0.5cm}0.61 &   \hspace{0.5cm}0.17 & \hspace{0.5cm}0.38 & \hspace{0.5cm} 0.92\\
            SET1c &\hspace{0.5cm} 0.81 &\hspace{0.5cm} 0.71 &\hspace{0.5cm}0.74 &  \hspace{0.4cm} 0.52 &\hspace{0.5cm}0.40 & \hspace{0.5cm}  1.02  \\
            SET1d &\hspace{0.5cm} -0.49 &\hspace{0.5cm} 0.33& \hspace{0.5cm}0.54 & \hspace{0.4cm}  0.44  &\hspace{0.5cm}0.17 &\hspace{0.5cm}  0.57 \\
            &&&&&&\\
            \hline
            \hline
            &&&&&&\\
            SET1e&\hspace{0.5cm} 0.34 &\hspace{0.5cm} 0.81 &\hspace{0.5cm} 0.89&\hspace{0.6cm}0.92  &\hspace{0.4cm} 0.43 &\hspace{0.5cm}  0.99\\       
            SET1f&\hspace{0.5cm} 0.54 &\hspace{0.5cm} 0.78&\hspace{0.5cm} 0.92&\hspace{0.6cm}3.38&\hspace{0.4cm} 0.55 &\hspace{0.5cm} 0.69 \\
            &&&&&&\\
            \hline
            \hline
            &&&&&&\\
            SET2a &\hspace{0.5cm} 1.57 &\hspace{0.5cm} 1.10 &\hspace{0.5cm}1.11 &\hspace{0.4cm} 47.45 &\hspace{0.5cm} 1.48 &\hspace{0.5cm}  33.7 \\     
            SET2b   &\hspace{0.5cm} 0.62&\hspace{0.5cm} 1.19&\hspace{0.5cm}1.91 &\hspace{0.4cm} 51.06&\hspace{0.5cm} 1.56 &\hspace{0.5cm}  33.9 \\ 
            SET2c  &\hspace{0.5cm} 1.09 &\hspace{0.5cm} 1.16&\hspace{0.5cm}1.90 &\hspace{0.4cm} 34.82 &\hspace{0.5cm} 1.43 &\hspace{0.5cm}  25.5 \\
            &&&&&&\\
            \hline
            \hline
            &&&&&&\\
            SET3a  &\hspace{0.8cm}-&\hspace{0.8cm}-&\hspace{0.8cm}- &\hspace{0.6cm}1.35 &\hspace{0.5cm}0.69&\hspace{0.5cm}  4 \\ 
            SET3b &\hspace{0.8cm}-&\hspace{0.8cm}-&\hspace{0.8cm}- &\hspace{0.5cm}  188.41 &\hspace{0.5cm} 5.93& \hspace{0.5cm}  123 \\  
            &&&&&&\\
           
    \hline
\end{tabular}
\end{centering}
\end{table*}


\begin{figure*}
\begin{multicols}{2}
\includegraphics[width=\linewidth]{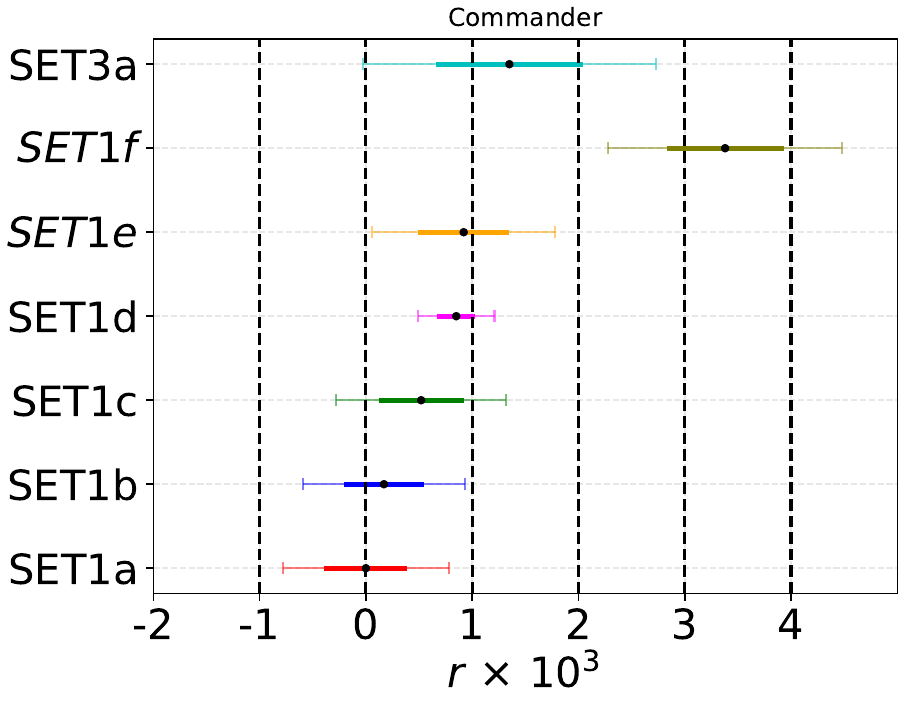} \par
\includegraphics[width=\linewidth]{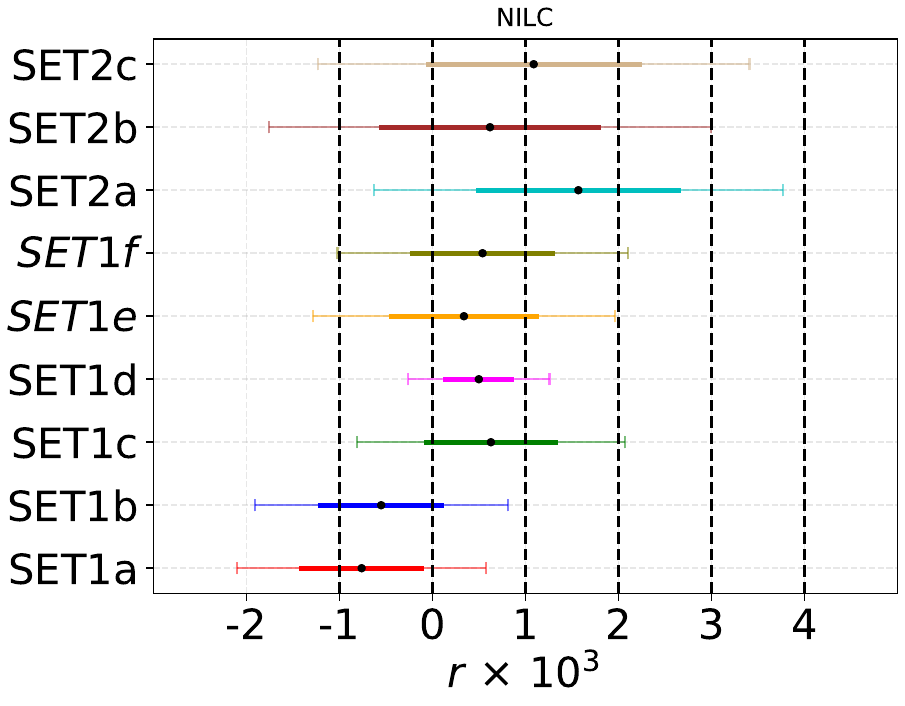}\par
\end{multicols}
\caption{The figure depicts the maximum likelihood values of $r$ (\textit{black dots}), 1$\sigma$ (\textit{thick lines}) and  2$\sigma$ (\textit{thin lines}) error bars for different sky configurations as listed in Table~\ref{table4} for \COMMANDER\ (\textit{left panel}) and \NILC\ (\textit{right panel}) pipelines.
Vertical \textit{black dashed} lines are drawn at $r = \{-1,0, 1, 2, 3, 4\} \times 10^{-3}$. The $r_{mp}$ values are consistent with null detection within their 2$\sigma$ limit for all sky configurations except for curved-power law synchrotron (SET1f) in case of \COMMANDER. Note that, we do not include the results for SET2a-c and SET3b for \COMMANDER\ as the foreground residuals are significantly high due to the mismatch between the data model we fit and the sky model in presence of dust decorrelation. The corresponding results are listed in Table~\ref{table5}.}
\label{fig1:snr}
\end{figure*}
\begin{figure*}
  \begin{center}
     \includegraphics[width=1.\textwidth]{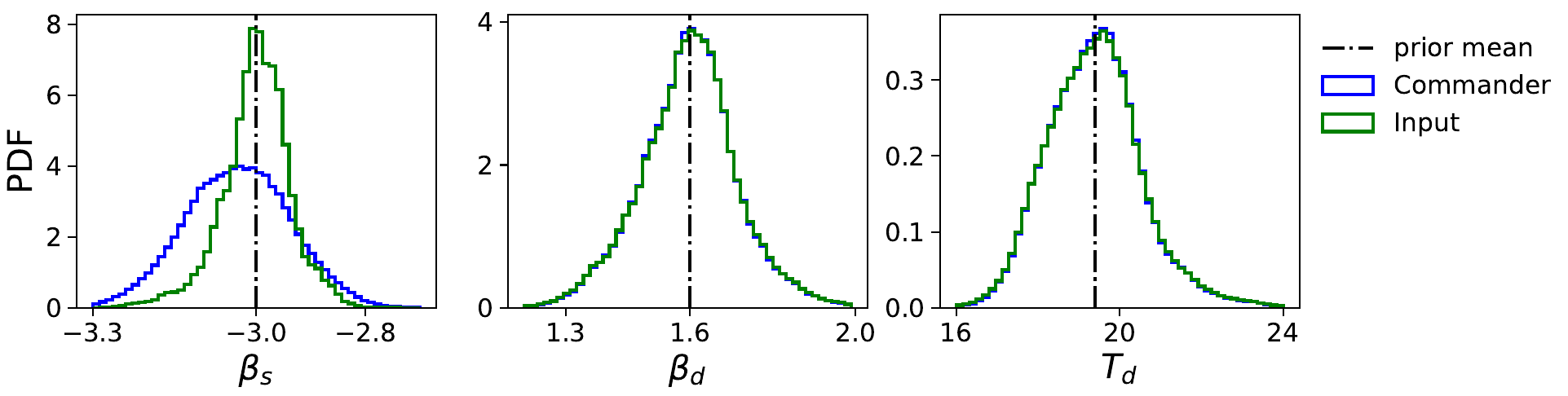}
     \end{center}
\caption{Comparison between probability density function (PDF) of \COMMANDER\ estimated foreground spectral parameters (\textit{blue}) and corresponding input spectral parameters (\textit{green}) for configuration SET1e. \textit{Left panel:} PDF of synchrotron spectral index $\beta_{s}$, \textit{Middle panel:} PDF of dust spectral index $\beta_d$ and \textit{Right panel:} PDF of thermal dust temperature $T_d$. The \textit{black dash-dotted} vertical lines are the corresponding mean values of Gaussian prior  adopted.} 
\label{fig1:inout_dist}
\end{figure*}


\begin{figure}
  \includegraphics[width=8.4cm]{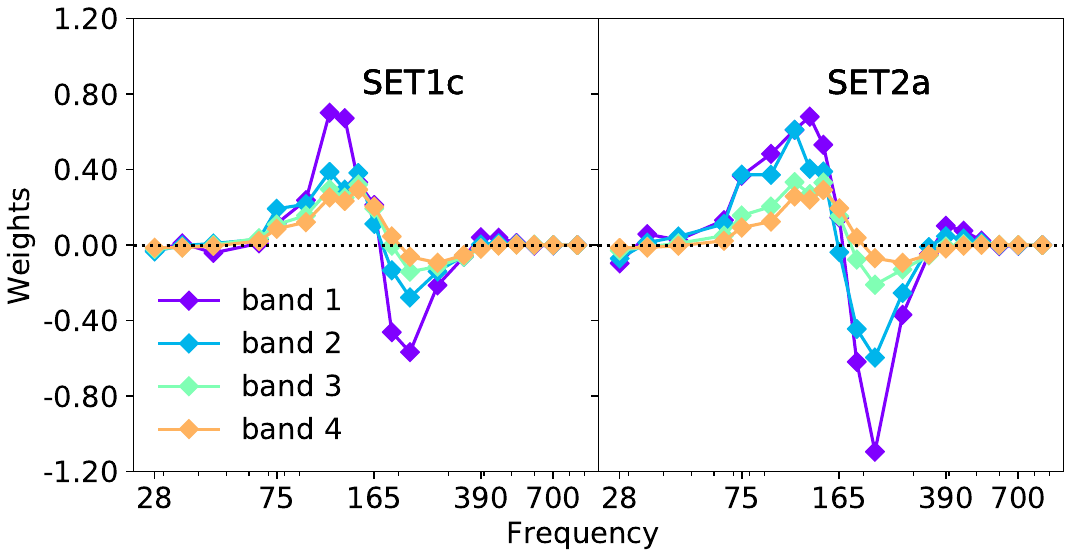}     

    \caption{Distribution of the full sky average values of the \NILC\ weights in the first four bands, across the \ECHO\ frequency channels for SET1c (\textit{left panel}) and SET2a (\textit{right panel}) sky configurations.} %
    \label{fig2:NILC_weights}
\end{figure}

\subsection{Map based results}
\label{sec:Map_based_results}
\subsubsection{Effect of sky coverage}
\label{sec:impact_of_mask}
Finding an optimal Galactic mask is very critical for our work as the level of the residuals of foreground for $B$-mode is not only high in the Galactic plane but also significant in the  upper and lower Galactic plane. We experiment with different sets of Galactic masks retaining sky fraction ranging from $90\%$ and to $20\%$ (see Fig.~\ref{fig:commander_nilc_mask}). Fig.~\ref{fig:posterior_fsky} shows the impact of the different Galactic masks on the measurement of tensor-to-scalar ratio for the simulations without AME and point sources (SET1a). The results clearly show that the bias in the measurement of $r$ ratio is minimum for the mask that retains the sky fraction $70\%$ for \NILC\ and $50\%$ for \COMMANDER. We notice that optimum sky fraction remains same for \COMMANDER\ while that reduces to 50\% for \NILC\ with addition of AME and point sources in simulation (SET1c).

\subsubsection{Impact of $\ell_{max}$ cut-off}
\label{sec:Robustness tests}

In Table~\ref{tab:lmaxcutoff}, we summarize the dependency of $r_{mp}$ and $\sigma (r)$ as a function of maximum $\ell$ cut-off ($\ell_{max}$) in the likelihood analysis for the sky configuration SET1a. Since most of the power for constraining $r$ is concentrated at $\ell \lesssim 120 - 150$, we find the $\sigma (r)$ improves consistently up to $\ell_{max} \sim$ 130. Further addition of power from the higher $\ell$ modes does not improve $\sigma (r)$. For our forecast results, we use the information of full range of available $\ell$ values: $2 \leq \ell \leq 180$ for \COMMANDER\ and $2 \leq \ell \leq 600$ for \NILC.


\subsubsection{Forecast results for baseline foreground model}
We begin our investigation with the results for our baseline sky configuration SET1a. In the foreground components, we only consider the polarized thermal dust with a single MBB spectrum and polarized synchrotron that follows the \galprop\ scaling. In Fig.~\ref{fig1:NILC_weights_set1a} we display the full sky average values of \NILC\ weights at 11 Needlet bands for all frequency channels. Inspecting  the \NILC\ weights, we find that most of the contribution in the final reconstruction of the CMB map comes from the frequency channels between 65 to 340 GHz. As foreground is the main contaminant at larger scales, the first four \NILC\ band weights largely drives the minimization of the foreground and changes significantly for different foreground models. At small angular scales, where noise is the dominant contaminant, we can see from the weights that even fewer frequency channels ($115-190$ GHz) contribute in the reconstruction of CMB. 

We estimate the $BB$ cross-power spectra over optimal masks (see Section~\ref{sec:impact_of_mask}) from the foreground cleaned two half-mission maps and use them in the likelihood analysis. In Fig.~\ref{fig1:examplecllikelihood}, we present the estimated $BB$ cross-power spectra from recovered CMB maps from \COMMANDER\ (left panel) and \NILC\ (right panel). The contributions of leakage from the foreground and noise to the cleaned CMB $B$-mode map are also shown for reference. Residual leakage has significant contribution in resulting power spectra mainly at low $\ell$ (< 60). 

The estimated $r$ bias are consistent with the null detection at the level of $1\sigma$. It implies that the residual foreground and noise does not introduce significant bias in the estimation of $r$ for the baseline simulation. 
 However, the foreground residuals inflate the associated uncertainties on $r$ to $0.79 \times 10^{-3}$ and $0.39 \times 10^{-3}$ for \NILC\ and \COMMANDER\ respectively which correspond to 75 \% and 49 \% increment as compared to the analytic forecast results with the idealised assumption of no foregrounds. The analytic forecast results rely on Fisher matrix formula: $\sigma (r) =  0.01\Bigg\{\sum_{\ell_{min}}^{\ell_{max}}\frac{(2\ell + 1)f_{sky}}{2}\Big[\frac{C_{\ell}^{BB,tensor} ( r = 0.01)}{C_{\ell}^{BB,lensing} +\, N^{BB}_{\ell}}\Big]^{2}\Bigg\}^{-1/2}$, where we consider \ECHO\ noise power spectra $N^{BB}_{\ell}$ as shown in Fig.~\ref{fig:echo_noise_spectra}. It implies that the uncertainty on the estimation of $r$ is mostly dominated by the foreground residuals even for the simplest assumptions about the polarized foreground emissions.

 Table~\ref{table5} summarizes the estimated values of maximum probable $r$ and 1$\sigma$ uncertainties for all the set of simulations considered in our analysis. o check for any possible bias present in the two component separation pipelines, we rerun the same analysis for independent set of CMB and noise realization for SET1a (keeping the foreground maps fixed). We report the maximum probable value of $r$ as $(-0.54 \pm 0.41) \times 10^{-3}$ (COMMANDER) and $(-0.15 \pm 0.82) \times 10^{-3}$  (NILC). These results imply that the two component separation pipelines are unbiased and the estimated maximum probable values of $r$ reported in Table~\ref{table5} is due to the statistical fluctuation for a particular CMB and noise realization.
 In order to check the goodness of fitting of power spectra, we also have listed $\chi^2$ per degrees of freedom (dof) in Table~\ref{table5}. In Fig.~\ref{fig1:snr}, we show a comparison of  $r_{mp}$ (black dots) along with 1$\sigma$ (thick lines) and 2$\sigma$ (thin lines) uncertainty for our baseline simulation with some other set of simulations.

\subsubsection{Effect of polarized AME and faint point sources}
\label{sec:impact_ame_ps}

The sky configurations SET1b and SET1c have been generated from the baseline configuration by adding respectively the polarized AME and AME, extragalactic faint point sources consistent with the current upper limits. Since \NILC\ is a blind component separation method, it adjusts the wights to minimize this extra complication of the sky. As far as the analysis of these sky using \COMMANDER\ is concerned, we have ignored the polarized AME and extragalactic point sources component while fitting the foreground model within the \COMMANDER\ pipeline. While the results show that the polarized AME component does not introduce any significant bias (consistent with null detection within 1$\sigma$) in the measurement of $r$ (see Table~\ref{table5} for reference), the extragalactic faint point sources exhibit a small bias (null detection within $\sim 2\sigma$) in the measure of $r$ even after applying very conservative mask of sky fraction $50\%$.

\subsubsection{Delensing assessment}
In our forecast, we explore the possibility of improvement of  sensitivity of $r$ measurement with one possible scenarios of delensing with 84 \% reduction of lensing power. We explore this scenario in sky configuration SET1d. In this configuration, we adopt the same foreground model as SET1a. 
The  estimated value is $r = (-0.49 \pm 0.33)\times 10^{-3}$ and $(0.44 \pm 0.17)\times 10^{-3}$ for \NILC\ and \COMMANDER\ respectively. The results show an improvement of $\sigma(r)$ by $\sim$ 50 \% as compared to the results for same foreground configuration without delensing in SET1a.

\subsubsection{Departures from baseline model: synchrotron and dust complexity}
\label{sec:dust_synchrotron_complexity}
In this section, we explore the set of sky configurations with different synchrotron emission laws and dust templates. We fix the model of polarized AME and point sources as used in SET1c. 

\textit{Synchrotron scaling} - We update the emission law of polarized synchrotron from  the \galprop\ scaling relation to the power-law and curved power-law behaviour in the sky configurations SET1e and SET1f respectively. For \NILC, the $r$ bias for the synchrotron power-law and curved power-law models are consistent with the \galprop\ scaling relation with a marginal increment of $\sigma (r)$. 
For both sets, and the values of $r_{mp}$ are consistent with the null detection within 2$\sigma$.

In \COMMANDER\ parameterization, we fit synchrotron by same spectral scaling as used in these two respective sky configurations. We find that $\sigma (r)$ increases marginally in comparison to same for \galprop\ scaling. Furthermore, both the scaling introduce significantly larger bias for \COMMANDER\ as compared to \NILC. For power-law scaling, although bias is consistent with zero within 3$\sigma$ limit, for curved power-law scaling, the bias is too large to be consistent with zero within  3$\sigma$. This additional bias on $r$ for \COMMANDER\ is attributed to lack of frequency channels in low frequency side of the spectrum. In Fig.~\ref{fig1:inout_dist}, we compare the distribution of input and \COMMANDER\ recovered maps of synchrotron spectral index, $\beta_s$ ({left panel}), dust spectral index, $\beta_d$ ({middle panel}), and dust temperature, $T_d$ ({right panel}) used in simulation in SET1e.
We find \COMMANDER\ fits thermal dust spectral parameters $T_d$ and $\beta_d$ with desired accuracy. However, lack of frequency channels $<$ 28 \GHz\ prevents \COMMANDER\ to fit synchrotron spectral index adequately. This incorrect fitting to the data at low frequency results in incorrect foreground subtraction and results in a large biased estimation of $r$.

\begin{figure}
  \includegraphics[width=8.4cm]{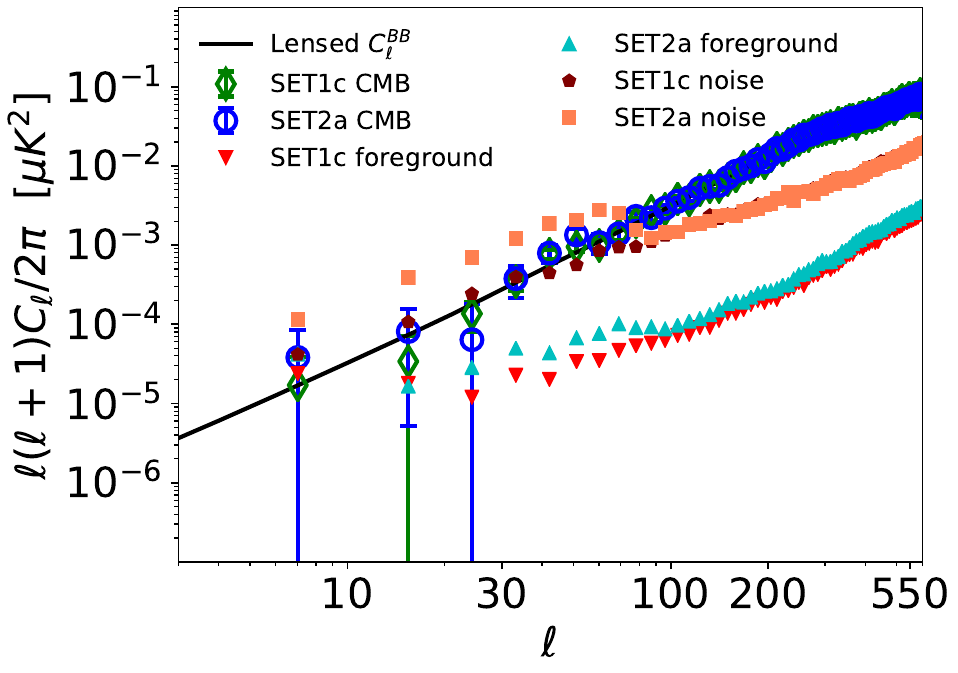}     
  \caption{Comparison of \NILC\ recovered CMB $BB$ power spectra for a single layer \GNILC\ dust model used in SET1c (\textit{green diamond}) and six-layers \GINES\ dust model used in SET2a (\textit{blue circle}). The other foreground components are same for both the sky configurations. The foreground and noise residuals are also compared for two simulations.}
    \label{fig2:NILC_aps_compare}
\end{figure}

\textit{Impact of decorrelation} - The frequency-frequency decorrelation of thermal dust may play a crucial role in introducing larger bias in $r$ measurement,  especially for parametric component separation. To investigate this, we first use three sky configurations SET2a-c where we use \GINES\ model that exhibits decorrelation across the frequencies. Three different synchrotron scaling are adopted in these three configurations. In Fig.~\ref{fig2:NILC_weights} we compare the \NILC\ weights for configurations SET1c and SET2a which exhibit simple dust and decorrelated dust model respectively. We find that to account for complexity in dust modelling the \NILC\ pipeline assigns higher weights to the first three Needlet bands. However, the level of residuals is still significantly higher as compared to configuration with simple dust model. This is clearly demonstrated in Fig.~\ref{fig2:NILC_aps_compare} where we compare the power spectra of residual noise and foreground for configurations SET1c and SET2a. The decorrelated thermal dust model in SET2a exhibits larger residual at large angular scales. As a result, the bias in $r$ increases by two-times and the uncertainty increases by $33\%$.

The forecast results for \COMMANDER\ show a very strong $r$ bias in comparison to the same for configurations in absence of dust decorrelation. This happens because the parametric component separation depend on whether the decorrelation across the \ECHO\ frequencies is properly parameterized in the dust model. Here, we consider one component MBB for fitting thermal dust in \COMMANDER\ that does not parameterize dust decorrelation. Because of incorrect assumption in dust modelling, we are effectively allowing the spectral mismatch between simulated dust in data and the model of the dust in parametric separation. Clearly this mismodelling of dust spectral property leaves a spurious dust residual in \COMMANDER\ recovered CMB maps resulting in a strong $r$ bias.


In order to investigate the impact of mismodelling in a more comprehensive manner, we consider two more simulations in configurations SET3a and SET3b where we introduce  \TDdust\ template. In SET3a, we do not introduce  decorrelation whereas  dust in SET3b exhibits decorrelation. We fit dust in both the configurations with one-component MBB spectrum. For SET3a, we find expected $r$ bias, since \COMMANDER\ does not suffer from mismodelling. However, for SET3b, we find a extremely large bias due to decorrelated dust, being attempted to be fit by a simple one-component MBB dust model. This implies adequate dust subtraction in presence of decorrelation is not possible with this simplest approach in the version of  \COMMANDER\ used here. We require different parameterization or multi-layer dust model in parametric separation to mitigate decorrelated dust.

We report the quality of fitting of model power spectrum to estimated $C_{\ell}$ in Table.~\ref{table5} measuring $\chi^2/d.o.f$. We find $\chi^2$/d.o.f values are close to one for \NILC\ pipeline except for the cases where we use decorrelated dust (SET2a-c). The decorrelated dust introduces excess bias at $\ell$ below 100 that prevents good fitting to the model power spectrum. For \COMMANDER\ the $\chi^2/d.o.f$ is close to one for the cases where there is no mismatch between data and foreground model (SET1a-e). Conversely, for SET1f, estimated $r$ has bias and $\chi^2/d.o.f$ is close to the baseline value because although the fitting to estimated $C_{\ell}$ is good, bias is due to the lack of frequency coverage of this experiment. The departure of $\chi^2$-values from baseline value for SET2a-c and 3b is due to the incorrect of dust model used in \COMMANDER. The estimated $r$ has bias and $\chi^2$ value is comparatively large for SET3a probably because we do not have access to the angular scales below $\ell$ = 40 for this simulation where the reionization peak is present.


\section{Conclusions}
\label{sec:Conclusions}
In this paper, we study whether instrument specification of \ECHO\ allows us efficient foreground subtraction to detect the primordial $B$-mode polarization in CMB to reach its scientific goal. 
Two component separation pipelines \NILC\ and \COMMANDER\ have been applied to set of foreground configurations covering a range of complexity over \ECHO\ channels followed by the power-spectrum estimation and likelihood analysis. This allows us to study ability of foreground subtraction of the pipelines for \ECHO\ instrument design through fully propagating the foreground residuals in estimation of $r$ sensitivity. Our simulations consist of dust and synchrotron with varying complexity, and two additional components due to polarized AME and point sources. The impact of gravitational lensing has also been tested comparing forecast results introducing fully lensed and 84\% delensed CMB maps.

If there is no spectral mismatch between simulated components in the data and parametric model fitting to the data, \COMMANDER\ can constraint $r$ for larger than $ 3 \times 10^{-3}$ at 3$\sigma$. However, we find need for more frequency channels at the low-frequency end of the spectrum to fit the power-law and curved power-law synchrotron. The blind component separation \NILC\ can constraint $r$  for larger than $ 2 \times 10^{-3}$ at 3$\sigma$. If 84\% delensing is possible to be achieved, we find the sensitivity can be improved by 50\% independent of component separation pipeline. The AME and point sources have a negligible impact on the sensitivity of $r$ measurement.

The \COMMANDER\ version used in this work cannot fit spectral parameters of dust adequately for the decorrelated dust model and hence subtracts foregrounds inadequately which results in immense bias on $r$. Therefore \ECHO\ cannot achieve its desired sensitivity with current \COMMANDER\ framework in presence of decorrelation. We argue that we should not ignore the modelling of decorrelation/multi-component dust in parametric foreground subtraction for such a high sensitive instrument. Since \NILC\ is a blind component separation, decorrelation does not deteriorate its performance much. However, decorrelation introduces larger residual leakage at large scale for \NILC\ that results in increment of $\sigma (r)$ by a factor of $\sim$ 1.5 in comparison to same in absence of decorrelation. 
We anticipate that impact of decorrelated dust may be possible to mitigate using some alternative approach of foreground separation, e.g. moment expansion based component separation algorithms \citep{chluba:2017, Ichiki:2019, Remazeilles:2020, Adak:2021, Rotti:2021}

A number of potential sources of systematic errors due to instrument imperfections and miscalibration are not considered in this paper. The treatment of those, and their possible interaction with the foreground cleaning algorithms, is left to future work. Another source of bias in our analysis comes from the Gaussian approximation of the likelihood function defined in Equation~\ref{eq:likelihood}. At low multipoles ($\ell < 30$), the likelihood function of the power spectrum is slightly non-Gaussian \citep{WMAP1:2003,Hamimeche:2008}. We will address the potential bias in $r$ estimation due to this effect in our forthcoming paper.

\section*{Acknowledgements}

DA acknowledges the University Grants Commission India for providing financial support as Senior Research Fellow. This work was supported by Science and Engineering Research Board, Department of Science and Technology, Govt. of India grant number SERB/ECR/2018/000826. All the computation using \COMMANDER\ code in this paper are done on the Pegasus cluster\footnote{\url{http://hpc.iucaa.in/}} at IUCAA. The results in this paper have been derived using the \healpix\ package. DA acknowledges Prof. Hans Kristian Eriksen and Dr. Ranajoy Banerji for helping to install \COMMANDER\ and preparing the parameter files. DA also acknowledges Dr. Shabbir Shaikh for useful discussion. AS acknowledges the use of Padmanabha cluster\footnote{\url{https://hpc.iisertvm.ac.in/}} at IISER-TVM for her work. AR acknowledges support from the ERC Consolidator Grant CMBSPEC (No. 725456) as part of the European Union’s Horizon 2020 research and innovation program.

\section*{Data Availability}
No external data are analysed in support of this research. The data will be shared on reasonable request to the corresponding author.

\bibliographystyle{mnras}
\bibliography{ECHO}

\begin{thebibliography}{}
\makeatletter
\relax
\def\mn@urlcharsother{\let\do\@makeother \do\$\do\&\do\#\do\^\do\_\do\%\do\~}
\def\mn@doi{\begingroup\mn@urlcharsother \@ifnextchar [ {\mn@doi@}
  {\mn@doi@[]}}
\def\mn@doi@[#1]#2{\def\@tempa{#1}\ifx\@tempa\@empty \href
  {http://dx.doi.org/#2} {doi:#2}\else \href {http://dx.doi.org/#2} {#1}\fi
  \endgroup}
\def\mn@eprint#1#2{\mn@eprint@#1:#2::\@nil}
\def\mn@eprint@arXiv#1{\href {http://arxiv.org/abs/#1} {{\tt arXiv:#1}}}
\def\mn@eprint@dblp#1{\href {http://dblp.uni-trier.de/rec/bibtex/#1.xml}
  {dblp:#1}}
\def\mn@eprint@#1:#2:#3:#4\@nil{\def\@tempa {#1}\def\@tempb {#2}\def\@tempc
  {#3}\ifx \@tempc \@empty \let \@tempc \@tempb \let \@tempb \@tempa \fi \ifx
  \@tempb \@empty \def\@tempb {arXiv}\fi \@ifundefined
  {mn@eprint@\@tempb}{\@tempb:\@tempc}{\expandafter \expandafter \csname
  mn@eprint@\@tempb\endcsname \expandafter{\@tempc}}}

\bibitem[\protect\citeauthoryear{Abazajian et~al.,}{Abazajian
  et~al.}{2022}]{CMBS4_2022}
Abazajian K.,  et~al., 2022, \mn@doi [The Astrophysical Journal]
  {10.3847/1538-4357/ac1596}, 926, 54

\bibitem[\protect\citeauthoryear{{Adak}}{{Adak}}{2021}]{Adak:2021}
{Adak} D.,  2021, \mn@doi [Monthly Notices of the Royal Astronomical Society]
  {10.1093/mnras/stab2392}, 507, 4618

\bibitem[\protect\citeauthoryear{{Adak}, {Ghosh}, {Boulanger}, {Haud},
  {Kalberla}, {Martin}, {Bracco}  \& {Souradeep}}{{Adak}
  et~al.}{2020}]{Adak:2019}
{Adak} D.,  {Ghosh} T.,  {Boulanger} F.,  {Haud} U.,  {Kalberla} P.,  {Martin}
  P.~G.,  {Bracco} A.,   {Souradeep} T.,  2020, \mn@doi [\aap]
  {10.1051/0004-6361/201936124}, \href
  {https://ui.adsabs.harvard.edu/abs/2020A&A...640A.100A} {640, A100}

\bibitem[\protect\citeauthoryear{Ade et~al.,}{Ade
  et~al.}{2014}]{Polarbear:2014}
Ade T. P. C. P. A.~R.,  et~al., 2014, \mn@doi [The Astrophysical Journal]
  {10.1088/0004-637x/794/2/171}, 794, 171

\bibitem[\protect\citeauthoryear{{Ade} et~al.,}{{Ade}
  et~al.}{2021}]{tristram2021planck_BK18}
{Ade} P.~A.~R.,  et~al., 2021, \mn@doi [\prl] {10.1103/PhysRevLett.127.151301},
  \href {https://ui.adsabs.harvard.edu/abs/2021PhRvL.127o1301A} {127, 151301}

\bibitem[\protect\citeauthoryear{Aghanim et~al.,}{Aghanim
  et~al.}{2020}]{aghanim2020planck}
Aghanim N.,  et~al., 2020, Astronomy \& Astrophysics, 641, A1

\bibitem[\protect\citeauthoryear{Aiola et~al.,}{Aiola
  et~al.}{2020}]{aiola2020atacama}
Aiola S.,  et~al., 2020, Journal of Cosmology and Astroparticle Physics, 2020,
  047

\bibitem[\protect\citeauthoryear{Akrami et~al.,}{Akrami
  et~al.}{2020}]{akrami2020planck}
Akrami Y.,  et~al., 2020, Astronomy \& Astrophysics, 641, A10

\bibitem[\protect\citeauthoryear{Albrecht \& Steinhardt}{Albrecht \&
  Steinhardt}{1982}]{PhysRevLett.48.1220}
Albrecht A.,  Steinhardt P.~J.,  1982, \mn@doi [Phys. Rev. Lett.]
  {10.1103/PhysRevLett.48.1220}, 48, 1220

\bibitem[\protect\citeauthoryear{Ali-Haïmoud, Hirata  \&
  Dickinson}{Ali-Haïmoud et~al.}{2009}]{Ali:2009}
Ali-Haïmoud Y.,  Hirata C.~M.,   Dickinson C.,  2009, \mn@doi [Monthly Notices
  of the Royal Astronomical Society] {10.1111/j.1365-2966.2009.14599.x}, 395,
  1055

\bibitem[\protect\citeauthoryear{Alonso, Dunkley, Thorne  \& N\ae{}ss}{Alonso
  et~al.}{2017}]{David:2017}
Alonso D.,  Dunkley J.,  Thorne B.,   N\ae{}ss S.,  2017, \mn@doi [Phys. Rev.
  D] {10.1103/PhysRevD.95.043504}, 95, 043504

\bibitem[\protect\citeauthoryear{{Alonso}, {Sanchez}, {Slosar}  \& {LSST Dark
  Energy Science Collaboration}}{{Alonso} et~al.}{2019}]{Namaster:2019}
{Alonso} D.,  {Sanchez} J.,  {Slosar} A.,   {LSST Dark Energy Science
  Collaboration} 2019, \mn@doi [\mnras] {10.1093/mnras/stz093}, \href
  {https://ui.adsabs.harvard.edu/abs/2019MNRAS.484.4127A} {484, 4127}

\bibitem[\protect\citeauthoryear{Andr{\'{e}} et~al.,}{Andr{\'{e}}
  et~al.}{2014}]{PRISM}
Andr{\'{e}} P.,  et~al., 2014, \mn@doi [Journal of Cosmology and Astroparticle
  Physics] {10.1088/1475-7516/2014/02/006}, 2014, 006

\bibitem[\protect\citeauthoryear{{Baleato Lizancos}, {Challinor}, {Sherwin}  \&
  {Namikawa}}{{Baleato Lizancos} et~al.}{2021}]{Baleato:2021}
{Baleato Lizancos} A.,  {Challinor} A.,  {Sherwin} B.~D.,   {Namikawa} T.,
  2021, arXiv e-prints, \href
  {https://ui.adsabs.harvard.edu/abs/2021arXiv210201045B} {p. arXiv:2102.01045}

\bibitem[\protect\citeauthoryear{Basak \& Delabrouille}{Basak \&
  Delabrouille}{2012}]{Basak_and_Delabrouille:2012}
Basak S.,  Delabrouille J.,  2012, \mn@doi [Monthly Notices of the Royal
  Astronomical Society] {10.1111/j.1365-2966.2011.19770.x}, 419, 1163

\bibitem[\protect\citeauthoryear{Basak \& Delabrouille}{Basak \&
  Delabrouille}{2013}]{Basak:2013}
Basak S.,  Delabrouille J.,  2013, \mn@doi [Monthly Notices of the Royal
  Astronomical Society] {10.1093/mnras/stt1158}, 435, 18

\bibitem[\protect\citeauthoryear{{Bennett} et~al.,}{{Bennett}
  et~al.}{2013}]{Bennett:2013}
{Bennett} C.~L.,  et~al., 2013, \mn@doi [\apjs] {10.1088/0067-0049/208/2/20},
  \href {https://ui.adsabs.harvard.edu/abs/2013ApJS..208...20B} {208, 20}

\bibitem[\protect\citeauthoryear{{Bonavera}, {Gonz{\'a}lez-Nuevo},
  {Arg{\"u}eso}  \& {Toffolatti}}{{Bonavera} et~al.}{2017}]{2017MNRAS4692401B}
{Bonavera} L.,  {Gonz{\'a}lez-Nuevo} J.,  {Arg{\"u}eso} F.,   {Toffolatti} L.,
  2017, \mn@doi [\mnras] {10.1093/mnras/stx1020}, \href
  {https://ui.adsabs.harvard.edu/abs/2017MNRAS.469.2401B} {469, 2401}

\bibitem[\protect\citeauthoryear{Bull et~al.,}{Bull et~al.}{2016}]{BULL201656}
Bull P.,  et~al., 2016, \mn@doi [Physics of the Dark Universe]
  {https://doi.org/10.1016/j.dark.2016.02.001}, 12, 56

\bibitem[\protect\citeauthoryear{{Chluba}, {Hill}  \& {Abitbol}}{{Chluba}
  et~al.}{2017}]{chluba:2017}
{Chluba} J.,  {Hill} J.~C.,   {Abitbol} M.~H.,  2017, \mn@doi [\mnras]
  {10.1093/mnras/stx1982}, \href
  {https://ui.adsabs.harvard.edu/abs/2017MNRAS.472.1195C} {472, 1195}

\bibitem[\protect\citeauthoryear{{Choi}}{{Choi}}{2020}]{ACT:2020}
{Choi} S. K. e.~a.,  2020, arXiv e-prints, \href
  {https://ui.adsabs.harvard.edu/abs/2020arXiv200707289C} {p. arXiv:2007.07289}

\bibitem[\protect\citeauthoryear{{Delabrouille, J.} et~al.,}{{Delabrouille, J.}
  et~al.}{2013}]{PSM:2011}
{Delabrouille, J.} et~al., 2013, \mn@doi [A\&A] {10.1051/0004-6361/201220019},
  553, A96

\bibitem[\protect\citeauthoryear{{Delabrouille}, {Cardoso}, {Le Jeune},
  {Betoule}, {Fay}  \& {Guilloux}}{{Delabrouille}
  et~al.}{2009}]{J.Delabrouille:2009}
{Delabrouille} J.,  {Cardoso} J.~F.,  {Le Jeune} M.,  {Betoule} M.,  {Fay} G.,
   {Guilloux} F.,  2009, \mn@doi [\aap] {10.1051/0004-6361:200810514}, \href
  {https://ui.adsabs.harvard.edu/abs/2009A&A...493..835D} {493, 835}

\bibitem[\protect\citeauthoryear{{Delabrouille} et~al.,}{{Delabrouille}
  et~al.}{2018}]{2018COREmission}
{Delabrouille} J.,  et~al., 2018, \mn@doi [JCAP]
  {10.1088/1475-7516/2018/04/014}, \href
  {https://ui.adsabs.harvard.edu/abs/2018JCAP...04..014D} {2018, 014}

\bibitem[\protect\citeauthoryear{{Delabrouille} et~al.,}{{Delabrouille}
  et~al.}{2019}]{Delabrouille1:2019}
{Delabrouille} J.,  et~al., 2019, arXiv e-prints, \href
  {https://ui.adsabs.harvard.edu/abs/2019arXiv190901591D} {p. arXiv:1909.01591}

\bibitem[\protect\citeauthoryear{Dickinson, Peel  \& Vidal}{Dickinson
  et~al.}{2011}]{Dickinson:2011}
Dickinson C.,  Peel M.,   Vidal M.,  2011, \mn@doi [Monthly Notices of the
  Royal Astronomical Society: Letters] {10.1111/j.1745-3933.2011.01138.x}, 418,
  L35

\bibitem[\protect\citeauthoryear{Dunne, Eales, Edmunds, Ivison, Alexander  \&
  Clements}{Dunne et~al.}{2000}]{Dunne:2000}
Dunne L.,  Eales S.,  Edmunds M.,  Ivison R.,  Alexander P.,   Clements D.~L.,
  2000, \mn@doi [Monthly Notices of the Royal Astronomical Society]
  {10.1046/j.1365-8711.2000.03386.x}, 315, 115

\bibitem[\protect\citeauthoryear{{Eriksen}, {Banday}, {G{\'o}rski}  \&
  {Lilje}}{{Eriksen} et~al.}{2004}]{Eriksen:2004}
{Eriksen} H.~K.,  {Banday} A.~J.,  {G{\'o}rski} K.~M.,   {Lilje} P.~B.,  2004,
  \mn@doi [\apj] {10.1086/422807}, \href
  {https://ui.adsabs.harvard.edu/abs/2004ApJ...612..633E} {612, 633}

\bibitem[\protect\citeauthoryear{{Eriksen}, {Jewell}, {Dickinson}, {Band ay},
  {G{\'o}rski}  \& {Lawrence}}{{Eriksen} et~al.}{2008}]{Eriksen:2008}
{Eriksen} H.~K.,  {Jewell} J.~B.,  {Dickinson} C.,  {Band ay} A.~J.,
  {G{\'o}rski} K.~M.,   {Lawrence} C.~R.,  2008, \mn@doi [\apj]
  {10.1086/525277}, \href
  {https://ui.adsabs.harvard.edu/abs/2008ApJ...676...10E} {676, 10}

\bibitem[\protect\citeauthoryear{{Errard}, {Feeney}, {Peiris}  \&
  {Jaffe}}{{Errard} et~al.}{2016}]{2016JCAP03052E}
{Errard} J.,  {Feeney} S.~M.,  {Peiris} H.~V.,   {Jaffe} A.~H.,  2016, \mn@doi
  [\jcap] {10.1088/1475-7516/2016/03/052}, \href
  {https://ui.adsabs.harvard.edu/abs/2016JCAP...03..052E} {2016, 052}

\bibitem[\protect\citeauthoryear{{Fauvet, L.} et~al.,}{{Fauvet, L.}
  et~al.}{2011}]{Fauvet:2011}
{Fauvet, L.} et~al., 2011, \mn@doi [A\&A] {10.1051/0004-6361/201014492}, 526,
  A145

\bibitem[\protect\citeauthoryear{{Finkbeiner}}{{Finkbeiner}}{2004}]{Finkbeiner:2004}
{Finkbeiner} D.~P.,  2004, \mn@doi [\apj] {10.1086/423482}, \href
  {https://ui.adsabs.harvard.edu/abs/2004ApJ...614..186F} {614, 186}

\bibitem[\protect\citeauthoryear{{Ghosh} et~al.,}{{Ghosh}
  et~al.}{2017}]{T_Ghosh:2017}
{Ghosh} T.,  et~al., 2017, \mn@doi [\aap] {10.1051/0004-6361/201629829}, \href
  {http://adsabs.harvard.edu/abs/2017A%26A...601A..71G} {601, A71}

\bibitem[\protect\citeauthoryear{{Green} et~al.,}{{Green}
  et~al.}{2015}]{Green:2015}
{Green} G.~M.,  et~al., 2015, \mn@doi [\apj] {10.1088/0004-637X/810/1/25},
  \href {https://ui.adsabs.harvard.edu/abs/2015ApJ...810...25G} {810, 25}

\bibitem[\protect\citeauthoryear{Gualtieri et~al.,}{Gualtieri
  et~al.}{2018}]{gualtieri2018spider}
Gualtieri R.,  et~al., 2018, Journal of Low Temperature Physics, 193, 1112

\bibitem[\protect\citeauthoryear{Guth}{Guth}{1981}]{Guth:1980}
Guth A.~H.,  1981, \mn@doi [Phys. Rev. D] {10.1103/PhysRevD.23.347}, 23, 347

\bibitem[\protect\citeauthoryear{Génova-Santos et~al.,}{Génova-Santos
  et~al.}{2015}]{Santos:2015}
Génova-Santos R.,  et~al., 2015, \mn@doi [Monthly Notices of the Royal
  Astronomical Society] {10.1093/mnras/stv1405}, 452, 4169

\bibitem[\protect\citeauthoryear{{Hamimeche} \& {Lewis}}{{Hamimeche} \&
  {Lewis}}{2008}]{Hamimeche:2008}
{Hamimeche} S.,  {Lewis} A.,  2008, \mn@doi [\prd]
  {10.1103/PhysRevD.77.103013}, \href
  {https://ui.adsabs.harvard.edu/abs/2008PhRvD..77j3013H} {77, 103013}

\bibitem[\protect\citeauthoryear{{Hanany} et~al.,}{{Hanany}
  et~al.}{2019}]{PICO:2019}
{Hanany} S.,  et~al., 2019, arXiv e-prints, \href
  {https://ui.adsabs.harvard.edu/abs/2019arXiv190210541H} {p. arXiv:1902.10541}

\bibitem[\protect\citeauthoryear{{Haslam}, {Salter}, {Stoffel}  \&
  {Wilson}}{{Haslam} et~al.}{1982}]{Haslam:1982}
{Haslam} C.~G.~T.,  {Salter} C.~J.,  {Stoffel} H.,   {Wilson} W.~E.,  1982,
  \aaps, \href {https://ui.adsabs.harvard.edu/abs/1982A&AS...47....1H} {47, 1}

\bibitem[\protect\citeauthoryear{Hazumi et~al.}{Hazumi
  et~al.}{2019}]{Hazumi:2019}
Hazumi M.,  et~al., 2019, \mn@doi [J. Low Temp. Phys.]
  {10.1007/s10909-019-02150-5}, 194, 443

\bibitem[\protect\citeauthoryear{Hivon, Gorski, Netterfield, Crill, Prunet  \&
  Hansen}{Hivon et~al.}{2002}]{Hivon:2002}
Hivon E.,  Gorski K.~M.,  Netterfield C.~B.,  Crill B.~P.,  Prunet S.,   Hansen
  F.,  2002, \mn@doi [The Astrophysical Journal] {10.1086/338126}, 567, 2

\bibitem[\protect\citeauthoryear{Hu \& White}{Hu \&
  White}{1997}]{Hu_and_white:1997}
Hu W.,  White M.,  1997, \mn@doi [New Astronomy]
  {https://doi.org/10.1016/S1384-1076(97)00022-5}, 2, 323

\bibitem[\protect\citeauthoryear{{Ichiki}, {Kanai}, {Katayama}  \&
  {Komatsu}}{{Ichiki} et~al.}{2019}]{Ichiki:2019}
{Ichiki} K.,  {Kanai} H.,  {Katayama} N.,   {Komatsu} E.,  2019, \mn@doi
  [Progress of Theoretical and Experimental Physics] {10.1093/ptep/ptz009},
  \href {https://ui.adsabs.harvard.edu/abs/2019PTEP.2019c3E01I} {2019, 033E01}

\bibitem[\protect\citeauthoryear{Jeffreys}{Jeffreys}{1939}]{Jeffreys:1939}
Jeffreys H.,  1939, {The Theory of Probability}.
Oxford Classic Texts in the Physical Sciences

\bibitem[\protect\citeauthoryear{{Kamionkowski} \& {Kosowsky}}{{Kamionkowski}
  \& {Kosowsky}}{1998}]{Kamionkowski:1998}
{Kamionkowski} M.,  {Kosowsky} A.,  1998, \mn@doi [\prd]
  {10.1103/PhysRevD.57.685}, \href
  {https://ui.adsabs.harvard.edu/abs/1998PhRvD..57..685K} {57, 685}

\bibitem[\protect\citeauthoryear{{Kazanas}}{{Kazanas}}{1980}]{1980ApJ...241L..59K}
{Kazanas} D.,  1980, \mn@doi [\apjl] {10.1086/183361}, \href
  {https://ui.adsabs.harvard.edu/abs/1980ApJ...241L..59K} {241, L59}

\bibitem[\protect\citeauthoryear{Knox \& Turner}{Knox \&
  Turner}{1993}]{Knox_Turner:1993}
Knox L.,  Turner M.~S.,  1993, \mn@doi [Phys. Rev. Lett.]
  {10.1103/PhysRevLett.70.371}, 70, 371

\bibitem[\protect\citeauthoryear{Kogut et~al.,}{Kogut
  et~al.}{2007}]{Kogut:2007}
Kogut A.,  et~al., 2007, \mn@doi [The Astrophysical Journal] {10.1086/519754},
  665, 355

\bibitem[\protect\citeauthoryear{{Krachmalnicoff, N.}, {Baccigalupi, C.},
  {Aumont, J.}, {Bersanelli, M.}  \& {Mennella, A.}}{{Krachmalnicoff, N.}
  et~al.}{2016}]{Krachmalnicoff:2016}
{Krachmalnicoff, N.} {Baccigalupi, C.} {Aumont, J.} {Bersanelli, M.}
  {Mennella, A.} 2016, \mn@doi [A\&A] {10.1051/0004-6361/201527678}, 588, A65

\bibitem[\protect\citeauthoryear{{Lazarian}}{{Lazarian}}{2007}]{Lazarian:2007}
{Lazarian} A.,  2007, \mn@doi [\jqsrt] {10.1016/j.jqsrt.2007.01.038}, \href
  {https://ui.adsabs.harvard.edu/abs/2007JQSRT.106..225L} {106, 225}

\bibitem[\protect\citeauthoryear{{Leitch}, {Readhead}, {Pearson}  \&
  {Myers}}{{Leitch} et~al.}{1997}]{Leitch:1997}
{Leitch} E.~M.,  {Readhead} A.~C.~S.,  {Pearson} T.~J.,   {Myers} S.~T.,  1997,
  \mn@doi [\apjl] {10.1086/310823}, \href
  {https://ui.adsabs.harvard.edu/abs/1997ApJ...486L..23L} {486, L23}

\bibitem[\protect\citeauthoryear{Linde}{Linde}{1982}]{linde1982new}
Linde A.~D.,  1982, Physics Letters B, 108, 389

\bibitem[\protect\citeauthoryear{Manzotti}{Manzotti}{2018}]{Manzotti:2017a}
Manzotti A.,  2018, \mn@doi [Phys. Rev. D] {10.1103/PhysRevD.97.043527}, 97,
  043527

\bibitem[\protect\citeauthoryear{{Manzotti} et~al.,}{{Manzotti}
  et~al.}{2017}]{Manzotti:2017}
{Manzotti} A.,  et~al., 2017, \mn@doi [\apj] {10.3847/1538-4357/aa82bb}, \href
  {https://ui.adsabs.harvard.edu/abs/2017ApJ...846...45M} {846, 45}

\bibitem[\protect\citeauthoryear{{Mart{\'\i}nez-Solaeche}, {Karakci}  \&
  {Delabrouille}}{{Mart{\'\i}nez-Solaeche} et~al.}{2018}]{gines:2018}
{Mart{\'\i}nez-Solaeche} G.,  {Karakci} A.,   {Delabrouille} J.,  2018, \mn@doi
  [\mnras] {10.1093/mnras/sty204}, \href
  {https://ui.adsabs.harvard.edu/abs/2018MNRAS.476.1310M} {476, 1310}

\bibitem[\protect\citeauthoryear{{Miville-Desch\^enes, M.-A.}, {Lagache, G.},
  {Boulanger, F.}  \& {Puget, J.-L.}}{{Miville-Desch\^enes, M.-A.}
  et~al.}{2007}]{Miville:2007}
{Miville-Desch\^enes, M.-A.} {Lagache, G.} {Boulanger, F.}  {Puget, J.-L.}
  2007, \mn@doi [A\&A] {10.1051/0004-6361:20066962}, 469, 595

\bibitem[\protect\citeauthoryear{{Miville-Desch{\^e}nes}, {Ysard}, {Lavabre},
  {Ponthieu}, {Mac{\'\i}as-P{\'e}rez}, {Aumont}  \&
  {Bernard}}{{Miville-Desch{\^e}nes} et~al.}{2008}]{Miville-Desch:2008}
{Miville-Desch{\^e}nes} M.~A.,  {Ysard} N.,  {Lavabre} A.,  {Ponthieu} N.,
  {Mac{\'\i}as-P{\'e}rez} J.~F.,  {Aumont} J.,   {Bernard} J.~P.,  2008,
  \mn@doi [\aap] {10.1051/0004-6361:200809484}, \href
  {https://ui.adsabs.harvard.edu/abs/2008A&A...490.1093M} {490, 1093}

\bibitem[\protect\citeauthoryear{{Moshir}, {Kopman}  \& {Conrow}}{{Moshir}
  et~al.}{1992}]{Moshir:1992}
{Moshir} M.,  {Kopman} G.,   {Conrow} T.~A.~O.,  1992, {IRAS Faint Source
  Survey, Explanatory supplement version 2}

\bibitem[\protect\citeauthoryear{{Mukherjee}, {Paul}  \&
  {Choudhury}}{{Mukherjee} et~al.}{2019}]{SuvodipMukherjee:2019}
{Mukherjee} S.,  {Paul} S.,   {Choudhury} T.~R.,  2019, \mn@doi [\mnras]
  {10.1093/mnras/stz1002}, \href
  {https://ui.adsabs.harvard.edu/abs/2019MNRAS.486.2042M} {486, 2042}

\bibitem[\protect\citeauthoryear{Narcowich, Petrushev  \& Ward}{Narcowich
  et~al.}{2006}]{doi:10.1137/040614359}
Narcowich F.~J.,  Petrushev P.,   Ward J.~D.,  2006, \mn@doi [SIAM Journal on
  Mathematical Analysis] {10.1137/040614359}, 38, 574

\bibitem[\protect\citeauthoryear{Orlando \& Strong}{Orlando \&
  Strong}{2013}]{Orlando:2013}
Orlando E.,  Strong A.,  2013, \mn@doi [Monthly Notices of the Royal
  Astronomical Society] {10.1093/mnras/stt1718}, 436, 2127

\bibitem[\protect\citeauthoryear{O’Dea, Clark, Contaldi  \&
  MacTavish}{O’Dea et~al.}{2011}]{ODea:2011}
O’Dea D.~T.,  Clark C.~N.,  Contaldi C.~R.,   MacTavish C.~J.,  2011, \mn@doi
  [Monthly Notices of the Royal Astronomical Society]
  {10.1111/j.1365-2966.2011.19851.x}, 419, 1795

\bibitem[\protect\citeauthoryear{{Pelgrims, V.}, {Clark, S. E.}, {Hensley, B.
  S.}, {Panopoulou, G. V.}, {Pavlidou, V.}, {Tassis, K.}, {Eriksen, H. K.}  \&
  {Wehus, I. K.}}{{Pelgrims, V.} et~al.}{2021}]{Pelgrims:2021}
{Pelgrims, V.} {Clark, S. E.} {Hensley, B. S.} {Panopoulou, G. V.} {Pavlidou,
  V.} {Tassis, K.} {Eriksen, H. K.}  {Wehus, I. K.} 2021, \mn@doi [A\&A]
  {10.1051/0004-6361/202040218}, 647, A16

\bibitem[\protect\citeauthoryear{{Planck Collaboration}}{{Planck
  Collaboration}}{2014}]{planck-XI:2013}
{Planck Collaboration} 2014, \mn@doi [A\&A] {10.1051/0004-6361/201323195}, 571,
  A11

\bibitem[\protect\citeauthoryear{{Planck Collaboration IV}}{{Planck
  Collaboration IV}}{2018}]{planck-IV:2018}
{Planck Collaboration IV} 2018, arXiv e-prints, \href
  {https://ui.adsabs.harvard.edu/abs/2018arXiv180706208P} {p. arXiv:1807.06208}

\bibitem[\protect\citeauthoryear{{Planck Collaboration VI}}{{Planck
  Collaboration VI}}{2018}]{planck-VI:2018}
{Planck Collaboration VI} 2018, arXiv e-prints, \href
  {https://ui.adsabs.harvard.edu/abs/2018arXiv180706209P} {p. arXiv:1807.06209}

\bibitem[\protect\citeauthoryear{{Planck Collaboration X}}{{Planck
  Collaboration X}}{2016}]{planck-x:2016}
{Planck Collaboration X} 2016, \mn@doi [\aap] {10.1051/0004-6361/201525967},
  \href {https://ui.adsabs.harvard.edu/abs/2016A&A...594A..10P} {594, A10}

\bibitem[\protect\citeauthoryear{{Planck Collaboration XI}}{{Planck
  Collaboration XI}}{2020}]{planck-XI:2018}
{Planck Collaboration XI} 2020, \mn@doi [\aap] {10.1051/0004-6361/201832618},
  \href {https://ui.adsabs.harvard.edu/abs/2020A&A...641A..11P} {641, A11}

\bibitem[\protect\citeauthoryear{{Planck Collaboration XLIV}}{{Planck
  Collaboration XLIV}}{2016}]{planck-XLIV:2016}
{Planck Collaboration XLIV} 2016, \mn@doi [\aap] {10.1051/0004-6361/201628636},
  \href {https://ui.adsabs.harvard.edu/abs/2016A&A...596A.105P} {596, A105}

\bibitem[\protect\citeauthoryear{{Planck Collaboration XLVIII}}{{Planck
  Collaboration XLVIII}}{2016}]{planck-XLVIII:2016}
{Planck Collaboration XLVIII} 2016, \mn@doi [\aap]
  {10.1051/0004-6361/201629022}, \href
  {https://ui.adsabs.harvard.edu/abs/2016A&A...596A.109P} {596, A109}

\bibitem[\protect\citeauthoryear{{Planck Collaboration XXV}}{{Planck
  Collaboration XXV}}{2016}]{planck-xxv:2015}
{Planck Collaboration XXV} 2016, \mn@doi [A\&A] {10.1051/0004-6361/201526803},
  594, A25

\bibitem[\protect\citeauthoryear{{Planck Collaboration} et~al.,}{{Planck
  Collaboration} et~al.}{2011}]{PEPI:2011}
{Planck Collaboration} et~al., 2011, \mn@doi [A\&A]
  {10.1051/0004-6361/201116464}, 536, A1

\bibitem[\protect\citeauthoryear{{Planck Collaboration} et~al.,}{{Planck
  Collaboration} et~al.}{2016}]{planck-XXVI:2016}
{Planck Collaboration} et~al., 2016, \mn@doi [\aap]
  {10.1051/0004-6361/201526914}, \href
  {https://ui.adsabs.harvard.edu/abs/2016A&A...594A..26P} {594, A26}

\bibitem[\protect\citeauthoryear{{Puglisi} et~al.,}{{Puglisi}
  et~al.}{2018}]{2018ApJ85885P}
{Puglisi} G.,  et~al., 2018, \mn@doi [\apj] {10.3847/1538-4357/aab3c7}, \href
  {https://ui.adsabs.harvard.edu/abs/2018ApJ...858...85P} {858, 85}

\bibitem[\protect\citeauthoryear{Remazeilles, Delabrouille  \&
  Cardoso}{Remazeilles et~al.}{2011}]{Remazeilles:2011}
Remazeilles M.,  Delabrouille J.,   Cardoso J.-F.,  2011, \mn@doi [Monthly
  Notices of the Royal Astronomical Society]
  {10.1111/j.1365-2966.2011.19497.x}, 418, 467

\bibitem[\protect\citeauthoryear{{Remazeilles}, {Dickinson}, {Eriksen}  \&
  {Wehus}}{{Remazeilles} et~al.}{2016}]{2016MNRAS4582032R}
{Remazeilles} M.,  {Dickinson} C.,  {Eriksen} H.~K.~K.,   {Wehus} I.~K.,  2016,
  \mn@doi [\mnras] {10.1093/mnras/stw441}, \href
  {https://ui.adsabs.harvard.edu/abs/2016MNRAS.458.2032R} {458, 2032}

\bibitem[\protect\citeauthoryear{{Remazeilles} et~al.,}{{Remazeilles}
  et~al.}{2018}]{CORE-B:2018}
{Remazeilles} M.,  et~al., 2018, \mn@doi [\jcap]
  {10.1088/1475-7516/2018/04/023}, \href
  {https://ui.adsabs.harvard.edu/abs/2018JCAP...04..023R} {2018, 023}

\bibitem[\protect\citeauthoryear{{Remazeilles}, {Rotti}  \&
  {Chluba}}{{Remazeilles} et~al.}{2020}]{Remazeilles:2020}
{Remazeilles} M.,  {Rotti} A.,   {Chluba} J.,  2020, arXiv e-prints, \href
  {https://ui.adsabs.harvard.edu/abs/2020arXiv200608628R} {p. arXiv:2006.08628}

\bibitem[\protect\citeauthoryear{{Ricci, R.}, {Prandoni, I.}, {Gruppioni, C.},
  {Sault, R. J.}  \& {De Zotti, G.}}{{Ricci, R.} et~al.}{2004}]{Ricci:2004}
{Ricci, R.} {Prandoni, I.} {Gruppioni, C.} {Sault, R. J.}  {De Zotti, G.} 2004,
  \mn@doi [A\&A] {10.1051/0004-6361:20034632}, 415, 549

\bibitem[\protect\citeauthoryear{{Rotti} \& {Chluba}}{{Rotti} \&
  {Chluba}}{2021}]{Rotti:2021}
{Rotti} A.,  {Chluba} J.,  2021, \mn@doi [\mnras] {10.1093/mnras/staa3292},
  \href {https://ui.adsabs.harvard.edu/abs/2021MNRAS.500..976R} {500, 976}

\bibitem[\protect\citeauthoryear{{Saikia} \& {Salter}}{{Saikia} \&
  {Salter}}{1988}]{Saikia:1988}
{Saikia} D.~J.,  {Salter} C.~J.,  1988, \mn@doi [\araa]
  {10.1146/annurev.aa.26.090188.000521}, \href
  {https://ui.adsabs.harvard.edu/abs/1988ARA&A..26...93S} {26, 93}

\bibitem[\protect\citeauthoryear{Sato}{Sato}{1981}]{10.1093/mnras/195.3.467}
Sato K.,  1981, \mn@doi [Monthly Notices of the Royal Astronomical Society]
  {10.1093/mnras/195.3.467}, 195, 467

\bibitem[\protect\citeauthoryear{Sayre et~al.,}{Sayre et~al.}{2020}]{SPT:2020}
Sayre J.~T.,  et~al., 2020, \mn@doi [Phys. Rev. D]
  {10.1103/PhysRevD.101.122003}, 101, 122003

\bibitem[\protect\citeauthoryear{Seljak \& Hirata}{Seljak \&
  Hirata}{2004}]{PhysRevD.69.043005}
Seljak U. c.~v.,  Hirata C.~M.,  2004, \mn@doi [Phys. Rev. D]
  {10.1103/PhysRevD.69.043005}, 69, 043005

\bibitem[\protect\citeauthoryear{{Seljak} \& {Zaldarriaga}}{{Seljak} \&
  {Zaldarriaga}}{1997}]{Seljak:1997}
{Seljak} U.,  {Zaldarriaga} M.,  1997, \mn@doi [\prl]
  {10.1103/PhysRevLett.78.2054}, \href
  {https://ui.adsabs.harvard.edu/abs/1997PhRvL..78.2054S} {78, 2054}

\bibitem[\protect\citeauthoryear{{Sherwin} \& {Schmittfull}}{{Sherwin} \&
  {Schmittfull}}{2015}]{Sherwin_Schmittfull:2015}
{Sherwin} B.~D.,  {Schmittfull} M.,  2015, \mn@doi [\prd]
  {10.1103/PhysRevD.92.043005}, \href
  {https://ui.adsabs.harvard.edu/abs/2015PhRvD..92d3005S} {92, 043005}

\bibitem[\protect\citeauthoryear{Smith}{Smith}{2006}]{SMITH:2006}
Smith K.~M.,  2006, \mn@doi [New Astronomy Reviews]
  {https://doi.org/10.1016/j.newar.2006.09.015}, 50, 1025

\bibitem[\protect\citeauthoryear{Smith \& Ferraro}{Smith \&
  Ferraro}{2017}]{Smith:2017}
Smith K.~M.,  Ferraro S.,  2017, \mn@doi [Phys. Rev. Lett.]
  {10.1103/PhysRevLett.119.021301}, 119, 021301

\bibitem[\protect\citeauthoryear{Starobinsky}{Starobinsky}{1980}]{STAROBINSKY:1980}
Starobinsky A.,  1980, \mn@doi [Physics Letters B]
  {https://doi.org/10.1016/0370-2693(80)90670-X}, 91, 99

\bibitem[\protect\citeauthoryear{{Tegmark} \& {Efstathiou}}{{Tegmark} \&
  {Efstathiou}}{1996}]{1996MNRAS.281.1297T}
{Tegmark} M.,  {Efstathiou} G.,  1996, \mn@doi [\mnras]
  {10.1093/mnras/281.4.1297}, \href
  {https://ui.adsabs.harvard.edu/abs/1996MNRAS.281.1297T} {281, 1297}

\bibitem[\protect\citeauthoryear{{The COrE Collaboration} et~al.,}{{The COrE
  Collaboration} et~al.}{2011}]{2011arXiv1102.2181T}
{The COrE Collaboration} et~al., 2011, arXiv e-prints, \href
  {https://ui.adsabs.harvard.edu/abs/2011arXiv1102.2181T} {p. arXiv:1102.2181}

\bibitem[\protect\citeauthoryear{{Tristram}, {Mac{\'\i}as-P{\'e}rez}, {Renault}
   \& {Santos}}{{Tristram} et~al.}{2005}]{XSPECT:2005}
{Tristram} M.,  {Mac{\'\i}as-P{\'e}rez} J.~F.,  {Renault} C.,   {Santos} D.,
  2005, \mn@doi [\mnras] {10.1111/j.1365-2966.2005.08760.x}, \href
  {https://ui.adsabs.harvard.edu/abs/2005MNRAS.358..833T} {358, 833}

\bibitem[\protect\citeauthoryear{{Trombetti}, {Burigana}, {De Zotti},
  {Galluzzi}  \& {Massardi}}{{Trombetti} et~al.}{2018}]{2018A&A618A29T}
{Trombetti} T.,  {Burigana} C.,  {De Zotti} G.,  {Galluzzi} V.,   {Massardi}
  M.,  2018, \mn@doi [\aap] {10.1051/0004-6361/201732342}, \href
  {https://ui.adsabs.harvard.edu/abs/2018A&A...618A..29T} {618, A29}

\bibitem[\protect\citeauthoryear{{Verde} et~al.,}{{Verde}
  et~al.}{2003}]{WMAP1:2003}
{Verde} L.,  et~al., 2003, \mn@doi [\apjs] {10.1086/377335}, \href
  {https://ui.adsabs.harvard.edu/abs/2003ApJS..148..195V} {148, 195}

\bibitem[\protect\citeauthoryear{Wandelt, Larson  \& Lakshminarayanan}{Wandelt
  et~al.}{2004}]{Wandelt:2004}
Wandelt B.~D.,  Larson D.~L.,   Lakshminarayanan A.,  2004, \mn@doi [Phys. Rev.
  D] {10.1103/PhysRevD.70.083511}, 70, 083511

\bibitem[\protect\citeauthoryear{de Bernardis et~al.,}{de~Bernardis
  et~al.}{2018}]{core_instrument:2018}
de Bernardis P.,  et~al., 2018, \mn@doi [Journal of Cosmology and Astroparticle
  Physics] {10.1088/1475-7516/2018/04/015}, 2018, 015

\bibitem[\protect\citeauthoryear{{de Oliveira-Costa}, {Tegmark}, {Davies},
  {Guti{\'e}rrez}, {Lasenby}, {Rebolo}  \& {Watson}}{{de Oliveira-Costa}
  et~al.}{2004}]{Oliveira-Costa:2004}
{de Oliveira-Costa} A.,  {Tegmark} M.,  {Davies} R.~D.,  {Guti{\'e}rrez} C.~M.,
   {Lasenby} A.~N.,  {Rebolo} R.,   {Watson} R.~A.,  2004, \mn@doi [\apjl]
  {10.1086/421293}, \href
  {https://ui.adsabs.harvard.edu/abs/2004ApJ...606L..89D} {606, L89}

\makeatother
\end{thebibliography}


\appendix

\end{document}